\documentclass[pra]{revtex4-2}

\usepackage{amsmath}
\usepackage{graphicx}
\usepackage{hyperref}

\usepackage{dsfont}
\usepackage{microtype}
\usepackage{slashed}
\usepackage{xcolor}

\hypersetup{
  colorlinks,
  urlcolor={blue},
  linkcolor={blue},
  citecolor={blue}
}

\allowdisplaybreaks[1]

\newcommand{\centeredgraphics}[2][]{\vcenter{\hbox{\includegraphics[#1]{#2}}}}

\newcommand{\half}{\mbox{\small{$\frac{1}{2}$}}}
\newcommand{\MSbar}{\overline{\mbox{MS}}}
\newcommand{\order}{{\cal O}}

\begin{document}

\title{Anomalous dimensions and critical exponents for the Gross-Neveu-Yukawa 
model at five loops}
\author{J.A.~Gracey}
\email{gracey@liverpool.ac.uk}
\affiliation{Theoretical Physics Division, Department of Mathematical Sciences,
University of Liverpool, Liverpool, L69 3BX, United Kingdom}
\author{A.~Maier}
\email{amaier@ifae.es}
\affiliation{Institut de F\'{i}sica d’Altes Energies (IFAE), The Barcelona 
Institute of Science and Technology, Campus UAB, 08193 Bellaterra (Barcelona), 
Spain and \\
Grup de F\'{i}sica Te\`{o}rica, Dept. F\'{i}sica, Universitat Aut\`{o}noma de 
Barcelona, E-08193 Bellaterra, Barcelona, Spain}
\author{P.~Marquard}
\email{peter.marquard@desy.de}
\affiliation{Deutsches Elektronen–Synchrotron DESY, Platanenallee 6, 
15738 Zeuthen, Germany}
\author{Y.~Schr\"oder}
\email{yschroder@ubiobio.cl}
\affiliation{Centro de Ciencias Exactas, Departamento de Ciencias B\'{a}sicas, 
Universidad del B\'{i}o-B\'{i}o, Avenida Andr\'{e}s Bello 720, Chill\'{a}n, 
Chile}

\begin{frontmatter@abstract}
We renormalize the Gross-Neveu-Yukawa model with an $O(N)$ symmetry to 
$\order(\epsilon^5)$ in $d$~$=$~$4$~$-$~$\epsilon$ dimensions and determine the
anomalous dimensions of the fermion and scalar fields, $\beta$-functions as 
well as the scalar field's mass operator. These are used to construct several 
$N$ dependent critical exponents relevant for quantum transitions in 
semi-metals and in particular those connected with graphene in three dimensions
when $N$~$=$~$2$. Improved exponent estimates for scalar fermion transitions on
a honeycomb lattice, when $N$~$=$~$1$, as well as for $N$~$=$~$5$ are also
given to compare with results from other techniques such as the conformal 
bootstrap.
\end{frontmatter@abstract}

\maketitle


\section{Introduction}

Wilson's development of the renormalization group equation led to a beneficial
and practical tool to study critical phenomena with applications to phase
transitions seen in Nature, \cite{Wilson:1973jj}. In particular continuum 
analytic quantum field theory techniques could be used alongside numerical 
methods employed in discrete models to study critical points in the 
renormalization group flow. The ultimate goal being to estimate the observables
which are the critical exponents. If determined accurately enough the exponents
could be tested against experimental values and thereby confirm symmetry 
properties of the phase transition that are intrinsic to the underlying 
continuum or discrete field theory. 

One of the more widely known applications of the critical renormalization group 
equation is that of transitions in the three dimensional Heisenberg ferromagnet
where critical exponents are available at very high precision from techniques 
such as perturbation theory, Monte Carlo methods, functional renormalization 
group together with the recent incarnation of the conformal bootstrap 
summarized in \cite{Poland:2016chs} for instance. The latter method, for 
example, has generally produced exponents to the highest numerical precision. 
It is founded on exploiting the conformal symmetry inherent at a fixed point to
determine the scaling behaviour of field and operator correlation functions. 
See \cite{Henriksson:2022rnm}, for example, for a comprehensive overview of the 
current conformal field theory properties of the scalar field theories 
underlying the Heisenberg ferromagnet transition. 

One key principle of the study of critical phenomena is that seemingly 
different models used to understand, say the Heisenberg example, are related in
that they lie in the same universality class of the core fixed point. In this 
Heisenberg case the three dimensional fixed point is the Wilson-Fisher one, 
\cite{Wilson:1971vs}, corresponding to the non-trivial critical point of the 
underlying field theory. It is known that continuum scalar $\phi^4$ theory is 
one of the members of this universality class. Although its critical dimension 
is four it is still possible to derive exponent estimates in three dimensions. 
Indeed as the available loop order of $\phi^4$ theory renormalization group 
functions increase, \cite{Kleinert:1991rg,Kompaniets:2017yct,Schnetz:2016fhy}, 
there have been several resummation studies based on the $\epsilon$ expansion. 

However a second continuum field theory is believed to be present in the same
universality class which is the nonlinear $\sigma$ model whose critical
dimension is two. In fact there is in principle an infinite tower of theories
in even spacetime dimensions which reside in the same class. What they share
in common across all dimensions is the same core interaction with additional
spectator interactions included in specific dimensions to ensure
renormalizability and therefore calculability. This concept of tower is not
unrelated to the notion of ultraviolet completion.

Although this instance of connecting critical theories across dimensions to
extract precision estimates for observables is well-known it is not an isolated
case with physical applications. In more recent years a second main 
universality class has emerged which is connected to transitions in recently
developed materials. Perhaps the best known example is that of critical 
phenomena in graphene. This is a one atom thick sheet of carbon atoms connected
at the corners of a hexagonal lattice. Stretching a sheet of graphene can 
change its electrical properties and it is believed the critical properties of 
the transition from having semi-metal properties to becoming an insulator is
described by a resident theory of what is termed the Gross-Neveu (GN) or 
Gross-Neveu-Yukawa (GNY) universality class. The two main continuum field 
theories connected with the class are the two dimensional Gross-Neveu model, 
\cite{Gross:1974jv}, and the four dimensional Gross-Neveu-Yukawa theory, 
\cite{Zinn-Justin:1991ksq}. Both theories involve an $O(N)$ multiplet of Dirac
fermions coupled to a scalar field. In two dimensions the scalar is an 
auxiliary field, \cite{Gross:1974jv}, and the GN model reverts to a quartic 
fermion interaction. In the GNY model an extra quartic scalar interaction is 
required as a spectator interaction to ensure renormalizability. 

Unlike the purely scalar field theory universality class of the Heisenberg 
model the GNY class is more extensive. This is because the core scalar-Yukawa 
interaction can be adapted to include spin-related features as well as 
different fermion species. Moreover it is not unrelated to the Standard Model 
of particle physics which has the same underlying interactions but decorated 
with a more intricate general symmetry group. Indeed the GNY model offers 
itself as a potential laboratory to test possible beyond the standard model 
ideas. For instance, in \cite{Balents:1998hqb, Lee:2006if, Ponte:2012ru, 
Grover:2013rc} it was observed that for Lagrangians with certain field content 
critical points can emerge with symmetries that are not present in the original 
Lagrangian. In particular supersymmetry can actually emerge in a variety of 
theories within the broad GNY universality class at fixed points where the two 
critical couplings are equivalent. Moreover the scalar and fermion anomalous 
dimension exponents become equal at criticality. For example, see 
\cite{Fei:2016sgs} for an in-depth discussion.

Since the turn of the century there has been intense activity to produce
numerically precise exponent estimates for the GNY universality class with the 
graphene example of $N$~$=$~$2$ being the primary goal with the connection 
having been established in \cite{Herbut:2006cs,Herbut:2009vu}. Other values of 
$N$ are also studied as they relate to transitions in other materials. A 
variety of techniques have been applied such as Monte Carlo or lattice field 
theory, \cite{Karkkainen:1993ef, Chandrasekharan:2013aya, Li:2014aoa, 
He:2017sjp, Huffman:2017swn, Liu:2019xnb, Huffman:2019efk, Wang:2023tza}, 
functional renormalization group, \cite{Janssen:2014gea, Knorr:2016sfs,
Hawashin:2025ikp}, conformal bootstrap, \cite{Iliesiu:2017nrv, 
Erramilli:2022kgp, Mitchell:2024hix}, $1/N$ expansion, \cite{Gracey:1990wi, 
Gracey:1992cp, Vasiliev:1992wr, Vasiliev:1993pi, Gracey:1993kb, Gracey:1993kc},
and continuum perturbation theory, \cite{Gracey:2016mio, Mihaila:2017ble, 
Zerf:2017zqi, Ihrig:2018hho}. While different approximations will not always 
produce exact agreement of exponents, it is generally the case that the 
computational direction of travel in recent years is leading towards a 
consensual picture. Indeed the most recent conformal bootstrap analysis of 
\cite{Erramilli:2022kgp, Mitchell:2024hix} has produced the most accurate, with
respect to small uncertainties, numerical estimates for the graphene 
transition. 

One of the other main continuum field theory approaches is that of perturbation
theory with calculations to four loops being carried out in the GNY class in 
four dimensions in \cite{Zinn-Justin:1991ksq, Rosenstein:1993zf, 
Mihaila:2017ble, Zerf:2017zqi, Ihrig:2018hho} as well as in the generalization 
of that class recently recorded in \cite{Steudtner:2025blh}. Coupled with the 
parallel results to four loops in the GN model of \cite{Gross:1974jv, 
Wetzel:1984nw, Ludwig:1987rk, Gracey:1990sx, Luperini:1991sv, Gracey:1991vy, 
Gracey:2016mio} the authors of \cite{Ihrig:2018hho} produced a set of three 
dimensional exponent estimates based on the full four loop universality class 
data using several resummation techniques. 

With the subsequent advance made in conformal bootstrap technology, 
\cite{Erramilli:2022kgp, Mitchell:2024hix}, it is therefore the purpose of this
article to advance the perturbative information for the GNY theory to 
{\em five} loops. This is not a trivial exercise since it is beset with a 
significant increase in the number of Feynman graphs to compute at this order 
driven by the presence of two interactions. 

Having established all the relevant renormalization group functions at this
new order, the second phase of our investigation is to refine the four loop 
exponent estimates of \cite{Ihrig:2018hho}. We achieve this by employing two 
resummation methods that on the whole produce numerical values that are
generally mutually compatible with the more recent conformal bootstrap values. 
Our summation methods will allow us to construct an approximation to the 
$d$-dimensional structure of the respective exponents between two and four 
dimensions from which we extract their values in three dimensions. 

While the graphene application has clearly more general interest, we will 
consider other values of $N$. For instance in the GNY class the $N$~$=$~$1$ 
critical exponents describe the semi-metal to insulator transition for a 
spinless system. Although previous activity on this substance is not as 
extensive as that of graphene, we find a degree of convergence of the 
perturbative estimates as well as consensual agreement with other techniques. 
However in order to try and understand some of the subtleties of our 
resummation analysis the case of $N$~$=$~$5$ is also considered for comparison 
with other approaches. Indeed it appears that exponent estimates in the general
GNY class do agree more precisely as $N$ increases hinting that the situation 
with lower $N$ exponents would perhaps stabilize were higher loop order data 
available.

The article is organized as follows. We describe the salient aspects of the
underlying GNY theory Lagrangian in Section \ref{sectgny} including our
conventions and Feynman rules before summarizing the technicalities behind the
extension of previous results to five loops. The outcome of our computations
is recorded in Section \ref{sectres} where the two $\beta$-functions and field
anomalous dimensions are presented as well as the anomalous dimension of the 
scalar field mass operator. As a derivative goal of the five loop work is to 
establish critical exponents estimates for three dimensional materials we
discuss the derivation of the $\epsilon$ expansion for the relevant ones in
Section \ref{sectexpn2}. Section \ref{sectanalysis} is devoted to establishing
and analysing exponent estimates for several values of $N$ using two main 
resummation methods as well as a comparison of where these new values sit in 
relation to results from non-perturbation theory techniques. Concluding remarks
are provided in Section \ref{sectconc} ahead of Appendix \ref{applargen} and
\ref{appgnn2}. The former Appendix details the non-trivial consistency check of
the perturbative $\epsilon$ expansion of the exponents of Sections 
\ref{sectres} and \ref{sectexpn2} with the same expansion of the exponents 
evaluated in the large $N$ approximation. The latter Appendix records the four 
loop $\epsilon$ expansion of the GN model exponents in the same conventions 
used in this article.

\section{GNY model and method}
\label{sectgny}


We begin by recalling the structure of the basic GNY theory that we renormalize
at five loops. The Lagrangian in $d$~$=$~$4$~$-$~$\epsilon$ Euclidean spacetime
dimensions expressed in terms of renormalized quantities is
\begin{equation}
  \label{eq:L}
  \mathcal{L} = Z_\psi \bar{\psi}\slashed{\partial} \psi
  + Z_{\phi\bar{\psi}\psi} y \mu^{\frac{\epsilon}{2}} \phi \bar{\psi}\psi
  + \frac{1}{2} \phi(Z_{\phi^2} m^2 - Z_\phi \partial_i^2) \phi + Z_{\phi^4} \lambda \mu^\epsilon \phi^4,
\end{equation}
where $\phi$ is a real scalar and $\psi$ a multiplet of $N$ Dirac fermions. For
orientation with other perturbative computations in the GNY model we will 
follow the notation used in \cite{Zerf:2017zqi}. As we will employ dimensional 
regularization throughout the scale $\mu$ is introduced to ensure the coupling 
constants $y$ and $\lambda$ are dimensionless in the regularized Lagrangian. 
Aside from the underlying $O(N)$ symmetry of the fermion multiplet the 
Lagrangian possesses a discrete chiral symmetry
\begin{equation}
\psi ~\rightarrow \gamma^5 \psi ~~~,~~~
\bar{\psi} ~\rightarrow -~ \bar{\psi} \gamma^5 ~.
\label{chirsym}
\end{equation}
We use the short-hand notation $\slashed{q}$~$=$~$q_i \gamma_i$ with 
four-dimensional Euclidean Dirac matrices fulfilling
$\{ \gamma_i, \gamma_j \} = 2 \delta_{ij}\mathds{1}_4$. Treating the boson mass
$m$ as a small perturbation we obtain the following Feynman rules:
\begin{align}
  \label{psi_prop}
 Z_\psi\times \raisebox{-17pt}{\includegraphics{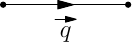}}\ ={}& -i\frac{\slashed{q}}{q^2} ~,\\[1em]
  \label{phi_prop}
 Z_\phi\times \raisebox{-17pt}{\includegraphics{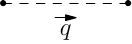}}\ ={}& \frac{1}{q^2} ~,\\[1em]
  \label{phi2_vx}
  \centeredgraphics{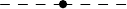}\ ={}& -Z_{\phi^2} m^2 ~,\\[1em]
  \label{phi4_vx}
  \centeredgraphics{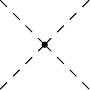}\ ={}& -4!  Z_{\phi^4} \lambda \mu^\epsilon ~,\\[1em]
  \label{psiphipsi_vx.pdf}
  \raisebox{-17pt}{\includegraphics{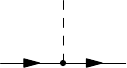}}\ ={}& -Z_{\phi\bar{\psi}\psi} g  \mu^{\frac{\epsilon}{2}} ~.
\end{align}
To determine the four renormalization constants
$Z_\psi$, $Z_\phi$, $Z_{\phi^4}$ and $Z_{\phi\bar{\psi}\psi}$ we respectively
consider the fermion and boson self-energies, the truncated four-boson vertex
and the truncated fermion-boson vertex. The remaining renormalization constant,
$Z_{\phi^2}$, is extracted from the $\mathcal{O}(m^2)$ contributions to the 
boson propagator, setting $m$~$=$~$0$ everywhere else. We use 
\texttt{QGRAF}~\cite{Nogueira:1991ex} to generate the Feynman diagrams to 
five-loop order.

The subsequent stage is to isolate the ultraviolet divergences by introducing 
an auxiliary mass 
$M$~\cite{Misiak:1994zw,vanRitbergen:1997va,Chetyrkin:1997fm}, which is 
independent of the boson mass $m$. This is achieved by adding a term
$\frac{1}{2} Z_{M^2} M^2 \phi^2$ to the Lagrangian in
equation~\eqref{eq:L}. Next, we turn the Feynman rules for the propagators 
$\mathcal{D}$, c.f.\ equations~\eqref{psi_prop} and \eqref{phi_prop}, into 
iterative equations of the form
\begin{equation}
  \label{eq:prop_iter}
  \mathcal{D} = \mathcal{D}_0 (1 - \mathcal{D}_0^{-1} \delta Z \mathcal{D}),
\end{equation}
where $\mathcal{D}_0$~$=$~$Z\mathcal{D}$ is the bare tree-level propagator and 
$\delta Z$~$=$~$Z$~$-$~$1$.  We then introduce the auxiliary mass $M$ in the 
denominator of the $\mathcal{D}_0$ prefactor. The resulting propagator Feynman 
rules read:
\begin{align}
  \label{psi_prop_reg}
 \mathcal{D}_\psi\equiv{}& \raisebox{-16pt}{\includegraphics{psi_prop.pdf}} = -i\frac{\slashed{q}}{q^2+M^2} (1-i\delta Z_\psi \slashed{q} \mathcal{D}_\psi),\\
  \label{phi_prop_reg}
 \mathcal{D}_\phi\equiv{}& \raisebox{-16pt}{\includegraphics{phi_prop.pdf}} = \frac{1}{q^2+M^2}(1 - q^2 \delta Z_\phi \mathcal{D}_\phi - M^2 Z_{M^2} \mathcal{D}_\phi).
\end{align}
With this we expand the fermion self-energy to linear order in the external
momentum and the $\mathcal{O}(m^0)$ contributions to the boson self-energy to 
quadratic order. In the $\mathcal{O}(m^2)$ contributions and in the vertices we
directly set the external momenta to zero.

With two mass scales, $m$ and $M$ one might, in principle, expect tadpole 
contributions to arise. While they are scaleless in the original formulation 
with $m$~$\to$~$0$, this is no longer the case after introducing the auxiliary 
mass $M$. However, the sum of degrees of all vertices forming a tadpole has to 
be odd, implying that the number of boson-fermion vertices is odd. Since all
fermion loops inside the tadpole are closed, at least one of them has an odd 
number of fermion propagators. As the auxiliary mass only occurs in the
denominators of the fermion propagators, the trace for this loop and therefore 
the whole tadpole vanishes.


Having summarized the algorithm that produces infrared regularized Feynman
graphs, we use ~\texttt{FORM}~\cite{Vermaseren:2000nd,Tentyukov:2007mu} to
insert the Feynman rules and expand the integrals in the external momentum. 
With the help of custom code~\cite{dynast} based on \texttt{nauty} and
\texttt{Traces}~\cite{mckay2013practical} we identify the resulting $3425333$ 
scalar integrals with four five-loop integral families of massive vacuum
diagrams, depicted in Figure~\ref{fig:tops}.

{\begin{figure}[hb]
\includegraphics[width=11.00cm,height=2.00cm]{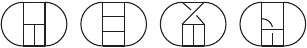}
\caption{There are four distinct fully massive five-loop vacuum-type integral 
sectors with $12$ propagators. All our integrals can be mapped onto these (or 
their sub-sectors that result after contracting a subset of propagator lines) 
after performing the infrared regularization as explained in the text.}
\label{fig:tops}
\end{figure}}

To reduce the scalar integrals to a small set of $110$ basis integrals we use 
Laporta's algorithm~\cite{Laporta:2001dd} as implemented in our custom codes 
\texttt{crusher}~\cite{crusher} and \texttt{tinbox}~\cite{tinbox}. Exploiting 
integration-by-parts identities~\cite{Chetyrkin:1981qh,Tkachov:1981wb}, we 
obtain large systems of linear equations, which we solve via finite-field
reconstruction techniques~\cite{Kauers:2008zz, Kant:2013vta, 
vonManteuffel:2014ixa, Peraro:2016wsq, Klappert:2019emp}.

To complete the reduction and map all results onto the minimal set of $110$ 
fully massive vacuum master integrals, we employ another in-house code 
{\texttt Spades} \cite{Luthe:2015ngq} which is again based on integration by
parts relations and the Laporta algorithm. This set of $5$-loop masters had 
been evaluated previously to very high numerical precision \cite{Luthe:2015ngq,
Luthe:2016sya, Luthe:2016spi}, using difference equations \cite{Laporta:2001dd}
and employing Fermat \cite{Fermat} for fast polynomial algebra. The 
corresponding numerical results satisfy a large number of internal consistency
checks, and have already contributed to various projects \cite{Luthe:2016ima, 
Luthe:2016xec, Luthe:2017ttc, Luthe:2017ttg} and hence can be regarded as
reliable. We note once more that, although our high-precision numerical results
exclude the $12$-propagator top-level families shown in Figure \ref{fig:tops}, 
for the renormalization constants those contribute only in three particular 
linear combinations. It had been observed before that those combinations can be
fixed by requiring consistent pole cancellations in renormalization as well as 
the absence of certain combinations of group factors, as explained in detail in
\cite{Luthe:2016xec}; see also \cite{Luthe:2017ttc}. The same mechanism is at 
play in our present calculation, such that we can finally employ PSLQ 
\cite{PSLQ} and express all our results in terms of zeta values only.

We close this section by recalling the connection of (\ref{eq:L}) with the
GN model of \cite{Gross:1974jv} which has the two dimensional Euclidean
Lagrangian
\begin{equation}
\mathcal{L}^{\mbox{\footnotesize{GN}}} ~=~ 
\bar{\psi}\slashed{\partial} \psi ~+~ h \phi \bar{\psi}\psi ~+~ 
\frac{1}{2} \phi^2
\label{laggn}
\end{equation}
where $h$ is the coupling constant. Renormalization constants have been
suppressed to avoid confusion with those of the four dimensional Lagrangian. 
The $\phi$ field now plays the role of an auxiliary field in two dimensions. 
Eliminating it from (\ref{laggn}) produces a quartic fermion interaction
leading to an asympotically free theory. However it is the shared scalar-Yukawa
interaction of (\ref{eq:L}) and (\ref{laggn}) that drives the Wilson-Fisher 
critical point equivalence of both theories between two and four dimensions. 
The GN Lagrangian also obeys the same discrete chiral symmetry as 
(\ref{chirsym}).

\section{Results}
\label{sectres}

Having described the field theory side of our computation to effect the full 
five loop renormalization we record the results of our mammoth calculation in
this section. The five renormalization constants are translated into the
renormalization group functions using standard methods and our expressions have
been determined in the $\MSbar$ scheme.

\subsection{$\beta$ functions}

First we record the two $\beta$-functions which are defined by
\begin{equation}
\beta_\lambda(\lambda,y) ~=~ \frac{d\lambda }{d \ln \mu} , \quad 
\beta_y(\lambda,y) ~=~ \frac{dy }{d \ln \mu} 
\end{equation}
and we will suppress the arguments $\lambda$ and $y$ throughout this section. 
To ease the presentation of the results we have chosen to give the coefficients
of each loop order separately by defining
\begin{equation}
  \beta_y = -\epsilon y + \sum_{L=1}^5 \beta_y^{(L)}, \quad   \beta_\lambda = -\epsilon \lambda + \sum_{L=1}^5 \beta_\lambda^{(L)}
\label{betaexp}
\end{equation}
where the $\epsilon$ terms reflect the dimensionlessness of the coupling
constants in the regularized theory. For $\beta_\lambda$ we have
\begin{eqnarray}
\beta_\lambda^{(1)}&=&36 \lambda ^2
+4 \lambda  N y
-N y^2 \nonumber \\
\beta_\lambda^{(2)}&=&-816 \lambda ^3
-72 \lambda ^2 N y
+7 \lambda  N y^2
+4 N y^3 \nonumber \\
%
\beta_\lambda^{(3)}&=&216 \left(96 \zeta _3+145\right) \lambda ^4
+1548 \lambda ^3 N y
-108 \lambda ^2 N^2 y^2
+\frac{3}{2} \left(648 \zeta _3+361\right) \lambda ^2 N y^2
+\frac{217}{2} \lambda  N^2 y^3  \nonumber \\
&&-\frac{3}{16} \left(624 \zeta _3+1465\right) \lambda  N y^3
-\frac{157}{8} N^2 y^4
+\left(\frac{5}{32}-12 \zeta _3\right) N y^4 \nonumber \\
%
\beta_\lambda^{(4)}&=&432 \left( 864 \zeta _4- \left(3744 \zeta _3+5760 \zeta _5+3499\right)\right) \lambda ^5
+36 \left(1728 \zeta _4-3456 \zeta _3-1355\right) \lambda ^4 N y
+24 \left(72 \zeta _3+263\right) \lambda ^3 N^2 y^2 \nonumber \\
&&-4 \left( 5184 \zeta _4+7452 \zeta _3+17280 \zeta _5+14521\right) \lambda ^3 N y^2
+144 \left(2 \zeta _3-1\right) \lambda ^2 N^3 y^3 \nonumber \\
&&+\left( 1944 \zeta _4-4392 \zeta _3-\frac{15649}{2}\right) \lambda ^2 N^2 y^3
+\left( 2511 \zeta _4+456 \zeta _3+10440 \zeta _5+\frac{211565}{16}\right) \lambda ^2 N y^3  \nonumber \\
&&+\left(\frac{1685}{12}-228 \zeta _3\right) \lambda  N^3 y^4
+\left(-675 \zeta _4+1039 \zeta _3-840 \zeta _5+\frac{17305}{12}\right) \lambda  N^2 y^4 \nonumber \\
&&+\left(-\frac{123}{2} \zeta _4+\frac{4677 \zeta _3}{4}+785 \zeta _5+\frac{9745}{16}\right) \lambda  N y^4
+\left(42 \zeta _3-\frac{193}{6}\right) N^3 y^5
+\left( 24 \zeta _4-45 \zeta _3+160 \zeta _5+\frac{1289}{12}\right) N^2 y^5 \nonumber \\
&&+\left(-\frac{231}{8} \zeta _4+\frac{277 \zeta _3}{4}+\frac{325 \zeta _5}{4}-\frac{4473}{128}\right) N y^5 \nonumber \\
%
\beta_\lambda^{(5)}&=&-108 \left(777600 \zeta _6+342432 \zeta _4- \left(103680 \zeta _3{}^2+1146960 \zeta _3+2274048 \zeta _5+3048192 \zeta _7+764621\right)\right) \lambda ^6 \nonumber \\
&&-18 \left( 691200 \zeta _6+ 404352 \zeta _4+ 110592 \zeta _3{}^2-499824 \zeta _3-1340928 \zeta _5-78653 \right) \lambda ^5 N y \nonumber \\
&&-\frac{27}{2} \left(23328 \zeta _4+ 8352 \zeta _3+4608 \zeta _5+29969 \right) \lambda ^4 N^2 y^2 
+\left( 3888 \zeta _4-22824 \zeta _3+9609\right) \lambda ^3 N^3 y^3 \nonumber \\
&&+\left( 3110400 \zeta _6+1560060\zeta _4+2363904 \zeta _3{}^2+3048516 \zeta _3+301968 \zeta _5+9344160 \zeta _7+\frac{19945539}{4}\right) \lambda ^4 N y^2 \nonumber \\
&&+\left(-259200 \zeta _6+ 27702 \zeta _4-62208 \zeta _3{}^2+378954 \zeta _3+662256 \zeta _5+\frac{993027}{2}\right) \lambda ^3 N^2 y^3 \nonumber \\
&&+\left(- 129600 \zeta _6- 8127 \zeta _4-67392 \zeta _3{}^2-414546 \zeta _3-115176 \zeta _5-1714608 \zeta _7-\frac{5584765}{8}\right) \lambda ^3 N y^3  \nonumber \\
&&+36 \left( 15 \zeta _4-10 \zeta _3-5\right) \lambda ^2 N^4 y^4 
+\left(- 7182 \zeta _4+18243 \zeta _3+3060 \zeta _5-\frac{124691}{8}\right) \lambda ^2 N^3 y^4 \nonumber \\
&&+\left( 106875 \zeta _6+\frac{3393 \zeta _4}{4}+36054 \zeta _3{}^2-\frac{88791 \zeta _3}{4}-165711 \zeta _5+77868 \zeta _7-\frac{8026377}{128}\right) \lambda ^2 N^2 y^4 \nonumber \\
&&+\left(\frac{82125 \zeta _6}{4}-\frac{218691 \zeta _4}{16}+\frac{9837 \zeta _3{}^2}{2}-\frac{505953 \zeta _3}{16}-57432 \zeta _5-\frac{969381 \zeta _7}{8}-\frac{13919569}{256}\right) \lambda ^2 N y^4 \nonumber \\
&&+\left(-450 \zeta _4+248 \zeta _3+\frac{699}{4}\right) \lambda  N^4 y^5 
+\left(-2100 \zeta _6+\frac{17379 \zeta _4}{8}-408 \zeta _3{}^2-\frac{20021 \zeta _3}{8}+3328 \zeta _5+\frac{1349851}{192}\right) \lambda  N^3 y^5 \nonumber \\
&&+\left(-\frac{44725 \zeta _6}{4}+\frac{12801 \zeta _4}{2}-\frac{1469 \zeta _3{}^2}{2}-9906 \zeta _3+\frac{24515 \zeta _5}{2}-16233 \zeta _7-\frac{14188555}{768}\right) \lambda  N^2 y^5 \nonumber \\
&&+\left( 1250 \zeta _6+\frac{30249 \zeta _4}{16}+\frac{1}{512} \left(-1533056 \zeta _3{}^2-4967520 \zeta _3-4893984 \zeta _5-2564016 \zeta _7+538157\right)\right) \lambda  N y^5 \nonumber \\
&&+\left(90 \zeta _4-\frac{157 \zeta _3}{4}-\frac{2623}{64}\right) N^4 y^6
+\left(\frac{1125 \zeta _6}{2}-\frac{603 \zeta _4}{4}+87 \zeta _3{}^2-\frac{1287 \zeta _3}{8}-899 \zeta _5-\frac{342865}{384}\right) N^3 y^6 \nonumber \\
&&+\left(\frac{18925 \zeta _6}{32}-\frac{25245 \zeta _4}{64}+\frac{-358464 \zeta _3{}^2-343920 \zeta _3-4913088 \zeta _5-1495872 \zeta _7+901799}{3072}\right) N^2 y^6 \nonumber \\
&&+\left(\frac{20075 \zeta _6}{64}+\frac{663 \zeta _4}{4}+\frac{-658048 \zeta _3{}^2+482720 \zeta _3-3539776 \zeta _5-1838592 \zeta _7+720901}{4096}\right) N y^6
\end{eqnarray}
where $\zeta_n$ is the Riemann zeta-function. Equally for $\beta_y$ we find
\begin{eqnarray}
\beta_y^{(1)}&=&2 N y^2
+3 y^2 \nonumber \\
%
\beta_y^{(2)}&=&24 \lambda ^2 y
-24 \lambda  y^2
-6 N y^3
-\frac{9 y^3}{8}  \nonumber \\
%
\beta_y^{(3)}&=&-216 \lambda ^3 y
-90 \lambda ^2 N y^2
+273 \lambda ^2 y^2
+90 \lambda  N y^3
+126 \lambda  y^3
+\frac{7 N^2 y^4}{2}
+\frac{1}{32} \left(432 \zeta _3+67\right) N y^4
+\frac{1}{64} \left(912 \zeta _3-697\right) y^4 \nonumber \\
%
\beta_y^{(4)}&=&14040 \lambda ^4 y
+288 \lambda ^3 N y^2
+36 \left(144 \zeta _3-455\right) \lambda ^3 y^2
-12 \lambda ^2 N^2 y^3
+2 \left(324 \zeta _3-635\right) \lambda ^2 N y^3
-\frac{135}{2} \left(40 \zeta _3+33\right) \lambda ^2 y^3 \nonumber \\
&&+12 \lambda  N^2 y^4
-\left(648 \zeta _3+683\right) \lambda N y^4
-\frac{3}{8} \left(1008 \zeta _3+943\right) \lambda  y^4
+\frac{11 N^3 y^5}{6}
+\left(\frac{27}{2} \zeta _4-\frac{125 \zeta _3}{2}-\frac{899}{24}\right) N^2 y^5 \nonumber \\
&&+\left(\frac{69}{2} \zeta _4-\frac{331 \zeta _3}{2}-105 \zeta _5+\frac{9907}{64}\right) N y^5
+\left(\frac{171}{8} \zeta _4+\frac{5 \zeta _3}{8}-\frac{215 \zeta _5}{2}+\frac{30529}{512}\right) y^5 \nonumber \\
\beta_y^{(5)}&=&-\frac{216}{5} \left(5760 \zeta _4-2160 \zeta _3+18545\right) \lambda ^5 y
-126 \left(72 \zeta _3+479\right) \lambda ^4 N y^2  \nonumber \\
&&+ 12 \left(19116 \zeta _4-16038 \zeta _3-34560 \zeta _5+77665\right) \lambda ^4 y^2
+162 \left(8 \zeta _3+9\right) \lambda ^3 N^2 y^3  \nonumber \\
&&+\left(8748 \zeta _4-55620 \zeta _3+84621\right) \lambda ^3 N y^3
+\left(-36450 \zeta _4+33822 \zeta _3+198720 \zeta _5+\frac{980055}{8}\right) \lambda ^3 y^3 \nonumber \\
&&+\frac{21}{2} \left(24 \zeta _3+5\right) \lambda ^2 N^3 y^4
+\left(1215 \zeta _4-3573 \zeta _3-\frac{3399}{2}\right) \lambda ^2 N^2 y^4 \nonumber \\
&&+\left(-4509 \zeta _4+\frac{96537 \zeta _3}{2}-15360 \zeta _5+\frac{463091}{16}\right) \lambda ^2 N y^4
+\left(-\frac{15255 \zeta _4}{4}+29601 \zeta _3+22725 \zeta _5+\frac{291657}{32}\right) \lambda ^2 y^4 \nonumber \\
&&-\frac{21}{2} \left(24 \zeta _3+5\right) \lambda  N^3 y^5
+\left(-1215 \zeta _4+3411 \zeta _3+3115\right) \lambda  N^2 y^5 \nonumber \\
&&+\left(-\frac{2007 \zeta _4}{2}+\frac{16629 \zeta _3}{2}+11100 \zeta _5-\frac{110555}{32}\right) \lambda  N y^5
-\frac{3}{64} \left(12720 \zeta _4-80768 \zeta _3-53280 \zeta _5+5603\right) \lambda  y^5 \nonumber \\
&&+\left(\frac{19}{16}-3 \zeta _3\right) N^4 y^6
+\left(-\frac{1125 \zeta _4}{16}+\frac{2707 \zeta _3}{16}+26 \zeta _5+\frac{3477}{128}\right) N^3 y^6 \nonumber \\
&&+\left(-\frac{1575 \zeta _6}{8}-\frac{6093 \zeta _4}{32}+\frac{273 \zeta _3{}^2}{4}+\frac{16345 \zeta _3}{32}+\frac{5259 \zeta _5}{4}-\frac{272751}{512}\right) N^2 y^6 \nonumber \\
&&+\left(-\frac{3975 \zeta _6}{8}-\frac{2439 \zeta _4}{8}+\frac{1}{512} \left(175296 \zeta _3{}^2-43816 \zeta _3+721360 \zeta _5+431928 \zeta _7-583115\right)\right) N y^6 \nonumber \\
&&-\frac{3 \left(7224000 \zeta _6+70560 \zeta _4-35 \left(191232 \zeta _3{}^2-326544 \zeta _3-22432 \zeta _5+582960 \zeta _7-107223\right)\right) y^6}{71680}
\end{eqnarray}

\subsection{Anomalous dimensions}
For the field anomalous dimensions we define 
\begin{equation}
\gamma_\psi = \sum_{L=1}^5 \gamma_\psi^{(L)}, 
\quad   
\gamma_\phi = \sum_{L=1}^5 \gamma_\phi^{(L)} 
\end{equation}
with
\begin{eqnarray}
\gamma_\psi^{(1)}&=&\frac{y}{2} \nonumber \\
%
\gamma_\psi^{(2)}&=&-\frac{3 N y^2}{4}
-\frac{y^2}{16} \nonumber \\
%
\gamma_\psi^{(3)}&=&-\frac{33 \lambda ^2 y}{2}
+6 \lambda  y^2
-\frac{3}{8} N^2 y^3
+\frac{47 N y^3}{32}
+\frac{3}{128} \left(16 \zeta _3-5\right) y^3 \nonumber \\
%
\gamma_\psi^{(4)}&=&342 \lambda ^3 y
+84 \lambda ^2 N y^2
+\left(54 \zeta _3-\frac{641}{4}\right) \lambda ^2 y^2
-\frac{87}{2} \lambda  N y^3
+\frac{3}{8} \left(32 \zeta _3-93\right) \lambda  y^3
+\left(\zeta _3-\frac{3}{16}\right) N^3 y^4 \nonumber \\
	&&-\frac{73}{24} N^2 y^4
+\left(-\frac{9}{8} \zeta _4-\frac{107 \zeta _3}{16}+\frac{835}{384}\right) N y^4
+\left(-\frac{15}{16} \zeta _4-\frac{9 \zeta _3}{4}-\frac{5 \zeta _5}{4}+\frac{4465}{1024}\right) y^4
\nonumber \\
\gamma_\psi^{(5)}&=&-18 \left(54 \zeta _3+935\right) \lambda ^4 y
-\frac{1971}{2} \lambda ^3 N y^2
+\left(243 \zeta _4-5589 \zeta _3+\frac{156147}{16}\right) \lambda ^3 y^2
+\left(9 \zeta _3+\frac{123}{8}\right) \lambda ^2 N^2 y^3 \nonumber \\
&&+\frac{1}{40} \left(2430 \zeta _4-43470 \zeta _3+50245\right) \lambda ^2 N y^3
+\left(\frac{3213 \zeta _4}{8}+522 \zeta _3-810 \zeta _5+\frac{12455}{16}\right) \lambda ^2 y^3
+\left(\frac{63}{8}-9 \zeta _3\right) \lambda  N^2 y^4 \nonumber \\
&& +\left(63 \zeta _4+\frac{1005 \zeta _3}{4}-105 \zeta _5+\frac{8717}{32}\right) \lambda  N y^4
+\frac{3 \left(20880 \zeta _4+95 \left(544 \zeta _3-800 \zeta _5+393\right)\right) \lambda  y^4}{1280} \nonumber \\
&& +\frac{3}{160} \left(80 \zeta _4-80 \zeta _3-5\right) N^4 y^5
+\frac{1}{40} \left(90 \zeta _4-50 \zeta _3-155\right) N^3 y^5
+\frac{1}{960} \left(855 \zeta _4+35805 \zeta _3-3120 \zeta _5+7930\right) N^2 y^5 \nonumber \\
&& +\left(\frac{425 \zeta _6}{32}-\frac{69 \zeta _4}{64}-\frac{5 \zeta _3{}^2}{16}+\frac{6789 \zeta _3}{128}+\frac{117 \zeta _5}{4}-\frac{492529}{6144}\right) N y^5 \nonumber \\
&& +\frac{\left(1736000 \zeta _6-715680 \zeta _4+35 \left(11008 \zeta _3{}^2-51600 \zeta _3+93664 \zeta _5+21168 \zeta _7-87335\right)\right) y^5}{143360}
\end{eqnarray}
and
\begin{eqnarray}
\gamma_\phi^{(1)}&=&2 N y \nonumber \\
%
\gamma_\phi^{(2)}&=&24 \lambda ^2
-\frac{5 N y^2}{2} \nonumber \\
%
\gamma_\phi^{(3)}&=&-216 \lambda ^3
-90 \lambda ^2 N y
+30 \lambda  N y^2
+\frac{25 N^2 y^3}{4}
+\frac{3}{32} \left(16 \zeta _3+7\right) N y^3 \nonumber \\
%
\gamma_\phi^{(4)}&=&14040 \lambda ^4
+288 \lambda ^3 N y
-12 \lambda ^2 N^2 y^2
+8 \left(81 \zeta _3-91\right) \lambda ^2 N y^2
-76 \lambda  N^2 y^3
+3 \left(16 \zeta _3-83\right) \lambda  N y^3 \nonumber \\
&&+\left(\frac{101}{24}-6 \zeta _3\right) N^3 y^4
+\left(-\frac{9}{2} \zeta _4-\frac{53 \zeta _3}{2}+\frac{211}{24}\right) N^2 y^4
-\frac{1}{8} \left(30 \zeta _4+123 \zeta _3+40 \zeta _5-145\right) N y^4 \nonumber \\
%
\gamma_\phi^{(5)}&=&-216 \left(1152 \zeta _4- 432 \zeta _3+ 3709\right) \lambda ^5
-126 \left(72 \zeta _3+479\right) \lambda ^4 N y
+162 \left(8 \zeta _3+9\right) \lambda ^3 N^2 y^2 \nonumber \\
&&-54 \left(90 \zeta _4+598 \zeta _3-751\right) \lambda ^3 N y^2
+\frac{21}{2} \left(24 \zeta _3+5\right) \lambda ^2 N^3 y^3
+\left(1215\zeta _4-2979 \zeta _3+\frac{4863}{4}\right) \lambda ^2 N^2 y^3 \nonumber \\
&&+\left(\frac{6993 \zeta _4}{2}-993 \zeta _3-5640 \zeta _5+\frac{167039}{16}\right) \lambda ^2 N y^3
-\frac{157}{2} \lambda  N^3 y^4
+\left(90 \zeta _4+933 \zeta _3-420 \zeta _5+\frac{6709}{8}\right) \lambda  N^2 y^4 \nonumber \\
&&+\left(\frac{1359 \zeta _4}{4}+\frac{1263 \zeta _3}{2}-\frac{1425 \zeta _5}{2}+\frac{43175}{64}\right) \lambda  N y^4
+\left(-9 \zeta _4+4 \zeta _3+\frac{27}{8}\right) N^4 y^5 \nonumber \\
&&+\left(-\frac{477 \zeta _4}{16}+\frac{1483 \zeta _3}{16}+2 \zeta _5+\frac{21271}{384}\right) N^3 y^5
+\left(\frac{425 \zeta _6}{8}-\frac{39 \zeta _4}{2}+\frac{\zeta _3{}^2}{4}+\frac{1071 \zeta _3}{4}+\frac{473 \zeta _5}{4}-\frac{520729}{1536}\right) N^2 y^5 \nonumber \\
&&+\frac{\left(347200 \zeta _6-131040 \zeta _4+7 \left(9472 \zeta _3{}^2-16992 \zeta _3+3 \left(40288 \zeta _5+7056 \zeta _7-42019\right)\right)\right) N y^5}{7168} ~.
\end{eqnarray}
Finally for the scalar mass operator we have 
\begin{eqnarray}
\gamma_{\phi^2}^{(1)}&=&-12 \lambda \nonumber \\
%
\gamma_{\phi^2}^{(2)}&=&144 \lambda ^2
+24 \lambda  N y
-2 N y^2 \nonumber \\
%
\gamma_{\phi^2}^{(3)}&=&-6264 \lambda ^3
-288 \lambda ^2 N y
+36 \lambda  N^2 y^2
-\frac{3}{2} \left(120 \zeta _3+11\right) \lambda  N y^2
-16 N^2 y^3
-12 \left(\zeta _3-3\right) N y^3 \nonumber \\
%
\gamma_{\phi^2}^{(4)}&=& 1728 \left(36 \zeta _4+18 \zeta _3+187\right) \lambda ^4
+36 \left(96 \zeta _3+313\right) \lambda ^3 N y
-96 \left(3 \zeta _3+11\right) \lambda ^2 N^2 y^2 \nonumber \\
&&+ 6 \left(216 \zeta _4+1440 \zeta _3+949\right) \lambda ^2 N y^2
+\left(48-96 \zeta _3\right) \lambda  N^3 y^3
+\left(-360 \zeta _4+504 \zeta _3+\frac{2427}{2}\right) \lambda  N^2 y^3 \nonumber \\
&&+\left(- 621 \zeta _4+960 \zeta _3+1080 \zeta _5-\frac{38967}{16}\right) \lambda  N y^3
+\left(36 \zeta _3-22\right) N^3 y^4 \nonumber \\
&&+\left( 27 \zeta _4+10 \zeta _3+140 \zeta _5-\frac{651}{4}\right) N^2 y^4
+\left(- 63 \zeta _4-84 \zeta _3+\frac{315 \zeta _5}{2}-\frac{1423}{32}\right) N y^4 \nonumber \\
%
\gamma_{\phi^2}^{(5)}&=&-108 \left( 86400 \zeta _6+ 39744 \zeta _4+ \left(-20736 \zeta _3{}^2+72048 \zeta _3+2304 \zeta _5+166267\right)\right) \lambda ^5 \nonumber \\
&&-18 \left(13392 \zeta _4+19392 \zeta _3+ 6912 \zeta _5+ 23237\right) \lambda ^4 N y
+\frac{9}{10} \left(10080 \zeta _4+6240 \zeta _3+92365\right) \lambda ^3 N^2 y^2 \nonumber \\
&&-\frac{3}{4} \left( 345600 \zeta _6+ 9648 \zeta _4+ 82944 \zeta _3{}^2+606960 \zeta _3+55296 \zeta _5+908231 \right) \lambda ^3 N y^2
-144 \left(\frac{9 \zeta _4}{2}-27 \zeta _3+11\right) \lambda ^2 N^3 y^3 \nonumber \\
&&+\frac{9}{4} \left(3600 \zeta _4-13216 \zeta _3-1440 \zeta _5-37163\right) \lambda ^2 N^2 y^3 \nonumber \\
&&+\left( 51300 \zeta _6+ 27594 \zeta _4-26568 \zeta _3{}^2-25200 \zeta _3-70668 \zeta _5+\frac{972023}{8}\right) \lambda ^2 N y^3 \nonumber \\
&&+\left(-180 \zeta _4+120 \zeta _3+60\right) \lambda  N^4 y^4
+\left( 378 \zeta _4-1569 \zeta _3-540 \zeta _5+\frac{19641}{8}\right) \lambda  N^3 y^4 \nonumber \\
&&+\left( 225 \zeta _6+\frac{10377 \zeta _4}{4}-2718 \zeta _3{}^2-\frac{8763 \zeta _3}{4}-19947 \zeta _5+\frac{1811931}{128}\right) \lambda  N^2 y^4 \nonumber \\
&&+\left(\frac{23925 \zeta _6}{4}+\frac{85545 \zeta _4}{16}-\frac{3}{256} \left(221312 \zeta _3{}^2-121232 \zeta _3+891904 \zeta _5+604128 \zeta _7-867809\right)\right) \lambda  N y^4 \nonumber \\
&&+4 \left(18 \zeta _4-10 \zeta _3-7\right) N^4 y^5
+\left(350 \zeta _6+ 96 \zeta _4+68 \zeta _3{}^2-66 \zeta _3-714 \zeta _5-\frac{57691}{96}\right) N^3 y^5 \nonumber \\
&&+\left( \frac{4725 \zeta _6}{4}-\frac{1701 \zeta _4}{4}-\frac{3 \zeta _3{}^2}{2}+\frac{1129 \zeta _3}{4}-\frac{1847 \zeta _5}{2}-1603 \zeta _7+\frac{1274651}{768}\right) N^2 y^5 \nonumber \\
&&+\left(\frac{14325 \zeta _6}{16}-\frac{141 \zeta _4}{16}+\frac{729 \zeta _3{}^2}{8}+\frac{4331 \zeta _3}{4}-\frac{7}{512} \left(3680 \zeta _5+125840 \zeta _7+25931\right)\right) N y^5 
\end{eqnarray}
where
\begin{equation}
\gamma_{\phi^2} = \sum_{L=1}^5 \gamma_{\phi^2}^{(L)} ~. 
\end{equation}

With such a colossal computation it is important to outline the independent
checks we carried out to ensure the veracity of the five loop contributions.

First the non-simple poles in $\epsilon$ of all the new five loop
renormalizations constants pass the check that they are all pre-determined by 
the simple poles of the lower loop companions due to the basic property of the 
renormalization group equation. 

A second check is established on the computation algorithm at five loops by 
realizing that setting $y$~$=$~$0$ in (\ref{eq:L}) produces a Lagrangian with 
$N$ free fermions and scalar $\phi^4$ theory. Therefore taking the same limit 
in the renormalization group functions the expressions for $\gamma_\phi$, 
$\gamma_{\phi^2}$ and $\beta_\lambda$ should be equivalent to the direct 
evaluation of these renormalization group functions to five loops given in, for
example, \cite{Dittes:1977aq, Gorishnii:1983gp, Kleinert:1991rg}. It is 
reassuring to note that this check is also satisfied after observing that 
$\lambda$ has to be rescaled by a factor that ensures the same normalization of
the quartic interaction is used in the comparison. 

The final check we have undertaken is to compare the critical exponents derived
from the renormalization group functions at the Wilson-Fisher fixed point with 
the same quantities but evaluated in the large $N$ expansion. The latter 
exponents are available to several orders in powers of $1/N$ with the relevant 
ones for this check computed in \cite{Gracey:1990wi, Vasiliev:1992wr, 
Vasiliev:1993pi, Gracey:1993kb, Gracey:1993kc, Gracey:2017fzu, 
Manashov:2017rrx}. At criticality the exponents depend on two variables which 
are $\epsilon$ and $N$. Carrying out a double Taylor expansion of the large $N$
exponents in powers of $\epsilon$ and $1/N$ produces expressions which overlap 
with $\order(\epsilon^5)$ perturbatively determined exponents when these are 
equally expanded in large $N$. Moreover this check provides a test of the five 
loop results for non-zero $y$. In summarizing this procedure we note that 
fuller details are provided for completeness in Appendix \ref{applargen} and 
record that full consistency was satisfyingly found.

\section{Critical exponents for $N$~$=$~$2$}
\label{sectexpn2}

Having constructed the full renormalization group equations for the GNY model
the next step is to derive the core critical exponents at the Wilson-Fisher
fixed point in $d$~$=$~$4$~$-$~$\epsilon$ dimensions. Therefore we have first
solved (\ref{betaexp}) for the critical couplings $y^\star$ and $\lambda^\star$ 
as power series in $\epsilon$ to $\order(\epsilon^5)$. These values are then
substituted into $\gamma_\psi(\lambda,y)$, $\gamma_\phi(\lambda,y)$ and
$\gamma_{\phi^2}(\lambda,y)$ to determine the respective $\epsilon$ expansions 
of $\eta_\psi(\epsilon)$, $\eta_\phi(\epsilon)$ and $\eta_{\phi^2}(\epsilon)$. 
We have carried this out for a general value of $N$. As $y$ and $\lambda$ 
appear in $\beta_y^{(1)}$ and $\beta_\lambda^{(1)}$ in a coupled way the 
critical couplings will involve the quantity $\sqrt{(4 N^2 + 132 N + 9)}$,
\cite{Karkkainen:1993ef}, which leads to cumbersome expressions when $N$ is 
arbitrary. Therefore we have provided them and results in \cite{arXiv}. However
as one of our main interests is the case of $N$~$=$~$2$ we will record the 
exponents for this value partly for our subsequent analysis but also because 
the $N$ dependent square root then reduces to an integer. For the two fields 
their anomalous dimensions are
\begin{equation}
\begin{split}
\eta_\psi = & ~
\frac{1}{14} \epsilon
- \frac{71}{10584} \epsilon^2
+ \bigg[
- \frac{2432695}{158696496}
- \frac{18}{2401} \zeta_3
\bigg]
\epsilon^3
+ \bigg[
\frac{150}{16807} \zeta_5
- \frac{27}{4802} \zeta_4
+ \frac{11109323}{555437736} \zeta_3
- \frac{111266497289}{11557548410688}
\bigg]
\epsilon^4
\\
& + \bigg[
- \frac{89279362217932877}{11783983899150199296}
- \frac{1905116200933}{377546581415808} \zeta_3
- \frac{136650473}{2744515872} \zeta_5
- \frac{5643}{537824} \zeta_7
- \frac{2004}{823543} \zeta_3^2
\\
& ~~~~~
+ \frac{375}{33614} \zeta_6
+ \frac{11109323}{740583648} \zeta_4
\bigg]
\epsilon^5 ~+~ \order(\epsilon^6)
\end{split}
\end{equation}
and
\begin{equation}
\begin{split}
\eta_\phi = & ~
\frac{4}{7} \epsilon
+ \frac{109}{882} \epsilon^2
+ \bigg[
\frac{1170245}{26449416}
- \frac{144}{2401} \zeta_3
\bigg]
\epsilon^3
+ \bigg[
\frac{20491307339}{481564517112}
- \frac{108}{2401} \zeta_4
+ \frac{1200}{16807} \zeta_5
+ \frac{1563532}{23143239} \zeta_3
\bigg]
\epsilon^4
\\
& + \bigg[
\frac{32079891787774525}{981998658262516608}
- \frac{1237285035017}{31462215117984} \zeta_3
- \frac{69575957}{228709656} \zeta_5
- \frac{15885}{823543} \zeta_3^2
- \frac{5643}{67228} \zeta_7
\\
& ~~~~~
+ \frac{1500}{16807} \zeta_6
+ \frac{390883}{7714413} \zeta_4
 \bigg]
\epsilon^5 ~+~ \order(\epsilon^6) ~.
\end{split}
\end{equation}
Equally the scalar field mass operator exponent is
\begin{equation}
\begin{split}
\eta_{\phi^2} = & ~
- \frac{8}{21} \epsilon
+ \frac{2942}{22491} \epsilon^2
+ \bigg[
\frac{88644929}{2547960408}
- \frac{144404}{1102059} \zeta_3
\bigg]
\epsilon^3
\\
& + \bigg[
\frac{18303692723219}{835032872672208}
- \frac{36101}{367353} \zeta_4
+ \frac{155735}{1361367} \zeta_5
+ \frac{1707335624}{20065188213} \zeta_3
\bigg]
\epsilon^4
\\
& + \bigg[
\frac{3531904209162553649}{212848209178400474784}
- \frac{6505866583}{16524272646} \zeta_5
+ \frac{24305439575}{327731407479} \zeta_3^2
+ \frac{44217078119105}{6819435126823032} \zeta_3
\\
& ~~~~~
+ \frac{778675}{5445468} \zeta_6
+ \frac{8001521}{13714512} \zeta_7
+ \frac{426833906}{6688396071} \zeta_4
\bigg]
\epsilon^5 ~+~ \order(\epsilon^6).
\end{split}
\end{equation}
While $\eta_{\phi^2}$ is an exponent that is not ordinarily determined
experimentally, it is required as an intermediate step to the one that is 
measured which is $1/\nu$ and derived from the scaling law 
\begin{equation}
\frac{1}{\nu} ~=~ 2 ~-~ \eta_\phi ~+~ \eta_{\phi^2} ~.
\end{equation}
Therefore we have 
\begin{equation}
\begin{split}
\frac{1}{\nu} = & ~
2 - \frac{20}{21} \epsilon
+ \frac{325}{44982}
\epsilon^2
+ \bigg[
- \frac{36133009}{3821940612}
- \frac{78308}{1102059} \zeta_3
\bigg]
\epsilon^3
\\
&
+ \bigg[
\frac{58535}{1361367} \zeta_5
- \frac{19577}{367353} \zeta_4
+ \frac{351753380}{20065188213} \zeta_3
- \frac{17228234202607}{835032872672208}
\bigg]
\epsilon^4
\\
& + \bigg[
- \frac{13685649343350298579}{851392836713601899136}
- \frac{5916014759}{66097090584} \zeta_5
+ \frac{30626922980}{327731407479} \zeta_3^2
+ \frac{1249594437836159}{27277740507292128} \zeta_3
\\
& ~~~~~
+ \frac{292675}{5445468} \zeta_6
+ \frac{9152693}{13714512} \zeta_7
+ \frac{87938345}{6688396071} \zeta_4
\bigg]
\epsilon^5 ~+~ \order(\epsilon^6) 
\end{split}
\end{equation}
which completes the set of core exponents that are required for our three 
dimensional analysis.

In addition we have determined the $\order(\epsilon^5)$ correction to scaling
exponents which are denoted by $\omega_\pm$ and are required as part of our
check with the large $N$ exponents. They are calculated from the gradients of 
the $\beta$-functions at criticality. In particular these are encoded in a 
$2$~$\times$~$2$ matrix
\begin{equation}
\beta_{ij} ~=~ \bigg(
\begin{array}{cc}
\frac{\partial \beta_y}{\partial y} & 
\frac{\partial \beta_y}{\partial \lambda} \\
\frac{\partial \beta_\lambda}{\partial y} & 
\frac{\partial \beta_\lambda}{\partial \lambda} \\
\end{array}
\bigg) ~\equiv~ 
\bigg(
\begin{array}{cc}
\beta_{11} & 
\beta_{12} \\
\beta_{21} & 
\beta_{22} \\
\end{array}
\bigg)
\end{equation}
which is then evaluated at $y^\star$ and $\lambda^\star$. The final stage is to
construct the $\epsilon$ expansion of the two perturbative eigenvalues of
$\beta_{ij}$ which correspond to $\omega_\pm$. In order to make a point of 
connection of our perturbative expressions for $\omega_\pm$ with the earlier 
large $N$ conventions of \cite{Manashov:2017rrx} we are required to take 
\begin{equation}
\omega_{\pm} = \frac{1}{2} \left( \beta_{11} + \beta_{22} \right) ~\mp~
\frac{1}{2} \sqrt{ \left[ (\beta_{11} - \beta_{22})^2 
+ 4 \beta_{12} \beta_{21} \right]}
\end{equation}
as the appropriate definitions. This leads to the two critical exponents
\begin{equation}
  \begin{split}
\omega_+ =
& ~ \frac{17}{7} \epsilon
- \frac{47725}{22491} \epsilon^2
+ \bigg[
+ \frac{25651541920}{8599366377}
+ \frac{1219592}{367353} \zeta_3
\bigg] \epsilon^3 \\
& + \bigg[
- \frac{45263181593447273}{7515295854049872}
- \frac{22241763116}{2866455459} \zeta_3
- \frac{2298610}{151263} \zeta_5
+ \frac{304898}{122451} \zeta_4
\bigg] \epsilon^4 \\
& + \bigg[
- \frac{5560440779}{955485153} \zeta_4
+ \frac{186794373160}{36414600831} \zeta_3^2
+ \frac{4733999225077}{99145635876} \zeta_5
+ \frac{253430775397933345}{13638870253646064} \zeta_3 \\
& ~~~~~
+ \frac{621984099876566907846887}{45975213182534502553344}
- \frac{5746525}{302526} \zeta_6
+ \frac{1467290683}{20571768} \zeta_7
\bigg] \epsilon^5 ~+~ \order(\epsilon^6)
  \end{split}
\label{peromegplu}
\end{equation}
and
\begin{equation}
  \begin{split}
\omega_- = 
& ~ \epsilon
~-~ \frac{533}{1512}\epsilon^2
~+~ \left[\frac{165\zeta_3}{686} + \frac{6685099}{34006392}\right]\epsilon^3 \\
& + \left[\frac{495\zeta_4}{2744} - \frac{61742201\zeta_3}{79348248} 
- \frac{1905\zeta_5}{4802} - \frac{11065294400875}{59438820397824}
\right] \epsilon^4 \\
& + \bigg[ - \frac{9525\zeta_6}{19208} 
- \frac{61742201\zeta_4}{105797664} + \frac{195609\zeta_3{}^2}{941192} 
+ \frac{60312023254291\zeta_3}{53935225916544} 
+ \frac{473424095\zeta_5}{196036848} + \frac{46467\zeta_7}{76832} \\
& ~~~~~+ \frac{10290966679190176397}{45452509325293625856} \bigg]\epsilon^5 ~+~
\order(\epsilon^6) ~.
  \end{split}
\label{peromegmin}
\end{equation}
The mismatch in the respective leading order coefficients in (\ref{peromegmin})
is due to the $\epsilon$ term of $\omega_-$ being $N$ independent unlike that 
of $\omega_+$. In relation to previous work we note that $\omega_-$ was denoted
by $\omega$ in \cite{Zerf:2017zqi} whereas $\omega_+$ corresponds to 
$\omega^\prime$ of the same article. Indeed it was noted in \cite{Zerf:2017zqi}
that the large $N$ expansion of the former perturbative eigen-exponent agreed 
with the $\epsilon$ dependence of the $\order(1/N)$ exponent directly 
determined in \cite{Gracey:2017fzu}. The large $N$ expressions for $\omega_\pm$
at $\order(1/N^2)$ appeared subsequently in \cite{Manashov:2017rrx}. 

\section{Analysis}
\label{sectanalysis}

Having constructed the five loop renormalization group functions our aim in 
this section is to produce refined estimates for critical exponents in three
dimensions for several values of $N$ that correspond to phase transitions in a 
variety of physical applications. For instance the critical exponents for 
$N$~$=$~$2$ relate to the semi-metal CDW (charge density wave) phase transition
of electrons in graphene while the same transition for the spinless fermions on
a honeycomb lattice is described by the $N$~$=$~$1$ GNY universality class. 
While both these cases are of physical interest we will also analyse 
$N$~$=$~$5$ since there are conformal bootstrap estimates available in 
\cite{Iliesiu:2017nrv}. We use this as a guidepost on the reliability of the 
resummation techniques we employ. Although the GN and GNY Lagrangians are 
renormalizable in two and four dimensions respectively since the models of 
physical interest are in three dimensions we have to translate the five loop 
$4$~$-$~$\epsilon$ exponents to three dimensions. Naively setting 
$\epsilon$~$=$~$1$ is problematic on several grounds. For instance when 
$N$~$=$~$2$ the numerical evaluation of the exponents gives
\begin{eqnarray}
\eta_\psi &=& 0.071429 \epsilon ~-~ 0.006708 \epsilon^2 ~-~
0.024341 \epsilon^3 ~+~ 0.017584 \epsilon^4 ~-~ 0.051782 \epsilon^5 ~+~
\order(\epsilon^6) \nonumber \\
\eta_\phi &=& 0.571429 \epsilon ~+~ 0.123583 \epsilon^2 ~-~
0.027849 \epsilon^3 ~+~ 0.149112 \epsilon^4 ~-~ 0.296922 \epsilon^5 ~+~
\order(\epsilon^6) \nonumber \\
\frac{1}{\nu} &=& 2 ~-~ 0.952381 \epsilon ~+~ 0.007225 \epsilon^2 ~-~
0.094868 \epsilon^3 ~-~ 0.012653 \epsilon^4 ~+~ 0.823067 \epsilon^5 ~+~
\order(\epsilon^6) 
\label{expd4numn2}
\end{eqnarray}
where it is evident that the $\order(\epsilon^5)$ terms are the same order of 
magnitude as the $\order(\epsilon)$ coefficients, unlike the intervening ones, 
meaning that the series convergence for $\epsilon$~$=$~$1$ could be 
questionable. To gauge the situation we have constructed a set of estimates 
using $[m/n]$ Pad\'{e} approximants which are recorded in Table \ref{pade} for
$N$~$=$~$2$. Pad\'{e} estimates were not possible for all choices at four 
loops. Cases where the approximant was not continuous from four dimensions to 
below three dimensions due, for instance, to poles in the range were excluded. 
The two continuous Pad\'{e} estimates at four loops are in reasonable accord. 
At five loops the $[1/4]$ values for $\eta_\psi$, $\eta_\phi$ and $\omega_-$ 
are significantly distant from the other three approximants which is probably 
related to the fact that they each begin with $\epsilon$. For both $\eta_\phi$ 
and $\omega_-$ the other five loop estimates are commensurate with the four 
loop ones. We note that a Monte Carlo estimate for $\omega_-$ of $0.8(1)$ was 
provided in \cite{Liu:2019xnb}. For $1/\nu$ there is more than a $10\%$ 
difference from the lowest to the highest value which could be a reflection 
that the five loop term of $1/\nu$ is around the same size as that at one loop 
with the intermediate values an order of magnitude or more smaller. For 
$\eta_\psi$ there appears to be a divergence of values for the three main 
approximants meaning that there is no real consensus for an estimate at five 
loops using this canonical Pad\'{e} analysis.

\begin{table}[ht]
\begin{tabular}{|c||c|c|c|c|}
\hline
  $~N ~=~ 2~$&$\eta_\psi$&$\eta_\phi$&$1/\nu$&$\omega_-$\\
\hline
  $[2/2]$&$0.0539$&$0.7079$&$0.931$&$0.794$\\
  $[3/1]$&$0.0506$&$0.6906$&$0.945$&$0.777$\\
\hline
  $[1/4]$&$~0.03457~$&$~0.4642~$&$~1.067~$&$~-~1.242~$\\
  $[2/3]$&$0.05065$&$0.7251$&$0.91917$&$0.7876$\\
  $[3/2]$&$0.02044$&$0.7291$&$1.039$&$0.7870$\\
  $[4/1]$&$0.04484$&$0.7170$&$0.9598$&$0.7882$\\
\hline
\end{tabular}
\caption{$L$ loop $[m/n]$ Pad\'{e} critical exponent estimates in $d$~$=$~$3$
where $L$~$=$~$m$~$+$~$n$.}
\label{pade}
\end{table}

In order to gauge where these naive Pad\'{e} and later estimates sit from an 
overall point of view we have provided a compendium of results from other 
methods in Table \ref{sumn2}. The entries are listed in chronological order 
beginning with large $N$ estimates where three terms of each exponent are known
in $d$-dimensions in a $1/N$ expansion. By this we mean $\eta_\phi$ and $1/\nu$
have been computed to $\order(1/N^2)$ as each canonical term is non-zero and 
$\eta_\psi$ is available at $\order(1/N^3)$ since there is no $\order(1)$ term.
The large $N$ estimates should be regarded as being within the same general
framework as perturbation theory which includes the three sets of four loop 
estimates from \cite{Ihrig:2018hho}. For $\eta_\psi$ there appears to be 
consistent agreement with results from more recent methods such as the 
conformal bootstrap results of \cite{Erramilli:2022kgp} and 
\cite{Mitchell:2024hix} which represent the highest precision to date with that
method. The four loop naive two dimensional Pad\'{e} estimates in Table 
\ref{sumn2} are the analogues of (\ref{expd4numn2}) and the estimates of Table 
\ref{pade} with the explicit numerical two dimensional $\epsilon$ expansions 
given in (\ref{expd2numn2}). The $[4/1]$ Pad\'{e} estimate of Table \ref{pade} 
is compatible with the four loop $\eta_\psi$ values of Table \ref{sumn2}. The 
various perturbative estimates for $\eta_\phi$ in Table \ref{sumn2} are not far
out of line with \cite{Erramilli:2022kgp} although the four loop approximants 
of Table \ref{sumn2} are around $2\%$ lower. For the $1/\nu$ Pad\'{e} estimates
the picture is not as concrete in relation to Table \ref{sumn2}. Given the 
degree of consistent estimates of the four loop two-sided Pad\'{e} and 
interpolating polynomial estimates with the most recent conformal bootstrap 
values of \cite{Erramilli:2022kgp} it would seem appropriate to extend that 
analysis with the new four dimensional four loop data.

{\begin{table}[ht]
\begin{center}
\begin{tabular}{|c||c|c|c|}
\hline
Method and source & $\eta_\psi$ & $\eta_\phi$ & $1/\nu$ \\
\hline
Large $N$ \cite{Gracey:1990wi,Gracey:1992cp,Vasiliev:1993pi,Vasiliev:1992wr,
Gracey:1993kb, Gracey:1993kc,Karkkainen:1993ef} & $0.044$ & $0.743$ & 
$0.952$ \\
Monte Carlo \cite{Karkkainen:1993ef} & ------ & $0.754(8)$ & $1.00(4)$ \\
Monte Carlo \cite{Chandrasekharan:2013aya} & $0.38(1)$ & $0.62(1)$ & 
$1.20(1)$ \\
Functional renormalization group \cite{Janssen:2014gea} & $0.032$ & $0.760$ & 
$0.982$ \\
Functional renormalization group \cite{Janssen:2014gea} & $0.033$ & $0.767$ & 
$0.978$ \\
Functional renormalization group \cite{Janssen:2014gea} & $0.032$ & $0.756$ & 
$0.982$ \\
Functional renormalization group \cite{Knorr:2016sfs} & $0.0276$ & $0.7765$ & 
$0.994(2)$ \\
Four loop $d$~$=$~$2$ naive Pad\'{e} \cite{Gracey:2016mio} & $0.082$ & 
$0.745$ & $0.931$ \\
Three loop $d$~$=$~$4$ naive Pad\'{e} \cite{Mihaila:2017ble} & $0.0740$ & 
$0.672$ & $1.048$ \\
Conformal bootstrap \cite{Iliesiu:2017nrv} & $0.044$ & $0.742$ & $0.880$ \\
Monte Carlo \cite{He:2017sjp} & ------ & $0.65(3)$ & $1.2(1)$ \\
Monte Carlo \cite{Huffman:2017swn} & ------ & $0.54(6)$ & $1.14(2)$ \\
Four loop $d$~$=$~$4$ naive Pad\'{e} \cite{Zerf:2017zqi} & $0.0539$ & 
$0.7079$ & $0.931$ \\
Four loop $d$~$=$~$4$ naive Pad\'{e} \cite{Zerf:2017zqi} & $0.0506$ & 
$0.6906$ & $0.945$ \\
Four loop two-sided Pad\'{e} \cite{Ihrig:2018hho} & $0.042$ & $0.735$ & 
$1.004$ \\
$~$Four loop interpolating polynomial \cite{Ihrig:2018hho}$~$ & $0.043$ & 
$0.731$ & $0.982$ \\
Four loop Pad\'{e}-Borel \cite{Ihrig:2018hho} & $0.043(12)$ & $0.704(15)$ &
$0.993(27)$ \\
Monte Carlo \cite{Liu:2019xnb} & $0.05(2)$ & $0.59(2)$ & $1.0(1)$ \\
Conformal bootstrap \cite{Erramilli:2022kgp} & $~0.04238(11)~$ &
$~0.7329(27)~$ & $~0.998(12)~$ \\
Monte Carlo \cite{Wang:2023tza} & $0.043(12)$ & $0.72(6)$ & 
$1.07(12)$ \\
Conformal bootstrap \cite{Mitchell:2024hix} & ------ & $0.7339(26)$ & 
$0.998(12)$ \\
Functional renormalization group \cite{Hawashin:2025ikp} & $0.032$ & $0.760$ & 
$0.982$ \\
\hline
\end{tabular}
\caption{Summary of previous exponent estimates for $N$~$=$~$2$.}
\label{sumn2}
\end{center}
\end{table}}

The basic idea behind the two-sided Pad\'{e} and interpolating polynomial
methods is that of ultraviolet completion. As the spacetime dimension of 
interest lies between two theories of the same universality class which are the
two dimensional GN and the four dimensional GNY models one can construct a 
function of $d$ such that the $\epsilon$ expansion of the function near two and
four dimensions equate to the $\epsilon$ expansion of one of the exponents 
$\eta_\psi$, $\eta_\phi$ and $1/\nu$. Once constructed the function of $d$ is 
evaluated for $d$~$=$~$3$. First we focus on the two-sided Pad\'{e} estimate 
which was used in \cite{Ihrig:2018hho} for instance. Its general definition is
\begin{equation}
{\cal P}_{[m/n]}(d) ~=~ 
\frac{\sum_{p=0}^m a_p d^p}{1 + \sum_{q=1}^n b_q d^q} 
\end{equation}
where the choice of unity for the denominator polynomial corresponds to a
normalization. This means that like the canonical Pad\'{e} approximant the 
number of possible estimates increases with loop order. What is different here 
is that there are more available data to determine the parameters $\{a_p\}$ and
$\{b_q\}$. In particular taking a generic $d$-dimensional exponent $\eta(d)$ 
its $\epsilon$ expansions around the respective critical dimensions are
\begin{equation}
\eta(2+\epsilon) ~=~ \sum_{n=0}^\infty \eta^{(2)}_n \epsilon^n ~~~,~~~
\eta(4-\epsilon) ~=~ \sum_{n=0}^\infty \eta^{(4)}_n \epsilon^n ~.
\label{genexp}
\end{equation}
where for instance $\eta^{(2)}_0$ and $\eta^{(2)}_1$ are both zero for
$\eta_\psi$. So amalgamating the known four loop GN model exponents with the 
five loop GNY ones means that $11$ parameters are available to find the 
$\{a_p\}$ and $\{b_q\}$ values. More generally if the exponent series for each 
critical dimension are known to $L$ loops then $2(L+1)$ coefficients are 
available leading in principle to $2(L+1)$ possible approximants which are
${\cal P}_{[m/2L+1-m]}(d)$ where $0$~$\leq$~$m$~$\leq$~$2L+1$. Like 
\cite{Ihrig:2018hho} we will follow the convention that only the diagonal 
approximants, ${\cal P}_{[L/L+1]}(d)$ and ${\cal P}_{[L+1/L]}(d)$, are reliable
and construct estimates for those. As noted in \cite{Ihrig:2018hho} continuity 
of each ${\cal P}_{[m/n]}(d)$ in $2$~$\leq$~$d$~$\leq$~$4$ is not guaranteed.

{\begin{table}[ht]
\begin{minipage}{7.0cm}
\centering
\begin{tabular}{|c||c|c|c|}
\hline
Pad\'{e} & $\eta_\psi$ & $\eta_\phi$ & $1/\nu$ \\
\hline
$~[1/2]~$ & ------ & $~0.786885~$ & $~1.012048~$  \\
$[2/1]$ & ------ & $0.812500$ & $1.000000$ \\
$[2/3]$ & ------ & $0.756156$ & ------ \\
$[3/2]$ & $~0.045930~$ & $0.752752$ & ------ \\
$[3/4]$ & $0.038789$ & $0.701041$ & ------ \\
$[4/3]$ & $0.040184$ & $0.727706$ & ------ \\
$[4/5]$ & $0.043732$ & $0.734702$ & $0.978327$ \\
$[5/4]$ & $0.041742$ & $0.734696$ & $1.004079$ \\
$[5/5]$ & $0.041418$ & $0.734530$ & $1.021288$ \\
\hline
\end{tabular}
\caption{Two-sided Pad\'{e} exponent estimates ${\cal P}_{[m/n]}(3)$ for 
$N$~$=$~$2$.}
\label{twopadn2}
\end{minipage}
\hspace{2.0cm}
\centering
\begin{minipage}{7.0cm}
\begin{tabular}{|c||c|c|c|}
\hline
Polynomial & $\eta_\psi$ & $\eta_\phi$ & $1/\nu$ \\
\hline
$~[1,1]~$ & $~0.017857~$ & $~0.809524~$ & $~1.011905~$ \\
$[1,2]$ & $0.034037$ & $0.774943$ & $1.025616$ \\
$[2,2]$ & $0.033636$ & $0.753047$ & $0.994951$ \\
$[2,3]$ & $0.036652$ & $0.746200$ & $1.003792$ \\
$[3,3]$ & $0.038991$ & $0.731170$ & $0.976204$ \\
$[3,4]$ & $0.040071$ & $0.739556$ & $0.987315$ \\
$[4,4]$ & $0.042592$ & $0.731413$ & $0.981959$ \\
$[4,5]$ & $0.041117$ & $0.730924$ & $1.010424$ \\
\hline
\end{tabular}
\caption{Interpolating polynomial ${\cal I}_{[i,j]}(3)$ exponent estimates for 
$N$~$=$~$2$.}
\label{intpolyn2}
\end{minipage}
\end{table}}

The second technique we will examine is that of using an interpolating 
polynomial \cite{Ihrig:2018hho}. Instead of constructing a $d$-dependent
rational polynomial a canonical polynomial is constructed where the 
coefficients are fixed from both $\epsilon$ expansions of a critical exponent. 
More particularly the polynomial is expressed in powers of $(d-2)$ so that the 
coefficients of the initial terms of the polynomial are given by those of the 
GN model. Then the higher order coefficients are determined from the $\epsilon$
expansion of the GNY theory in $d$~$=$~$4$~$-$~$\epsilon$ dimensions. 
Specifically
\begin{equation}
{\cal I}_{[i,j]}(d) ~=~ \sum_{m=0}^i \eta^{(2)}_m (d-2)^m ~+~ 
\sum_{n=i+1}^{i+j+1} c_n (d-2)^n 
\end{equation}
defines the interpolating polynomial \cite{Ihrig:2018hho} where the $\{c_n\}$ 
are deduced from the condition that expanding around $d$~$=$~$4$~$-$~$\epsilon$
reproduces the known coefficient of the same exponent. It is worth noting that
if instead ${\cal I}_{[i,j]}(d)$ was defined as a series in $(d-4)$ then the
{\em same} polynomial in $d$ would ultimately be produced. One advantage of 
using ${\cal I}_{[i,j]}(d)$ is that the function is continuous and cannot be 
singular unlike ${\cal P}_{[i/j]}(d)$. While this means a large number of 
estimates can be produced for each exponent from the ranges 
$1$~$\leq$~$i$~$\leq$~$4$ and $1$~$\leq$~$j$~$\leq$~$5$ we will focus on the 
estimates ${\cal I}_{[L,L]}(d)$ and ${\cal I}_{[L,L+1]}(d)$ at $L$ loops. The 
former is the natural choice as loop order increases but we include the latter 
in the analysis in order to see if any pattern is present when there is a 
mismatch in the number of terms available in the $\epsilon$ expansion in the 
respective critical dimensions. For both polynomial constructions we have 
solved for the respective unknown parameters in terms of the generic known 
coefficients of the respective series of (\ref{genexp}). This allowed us to 
construct procedures, as functions of the generic coefficients $\eta_n^{(2)}$ 
and $\eta_n^{(4)}$, in the computer algebra package {\tt Reduce}. Consequently 
evaluating the procedures with the explicit coefficients for a particular value
of $N$ allows us to quickly determine estimates. We chose not to construct 
explicit $N$-dependent expressions as they would be singular at 
$N$~$=$~$\half$.

{\begin{figure}[ht]
\includegraphics[width=8.50cm,height=7.00cm]{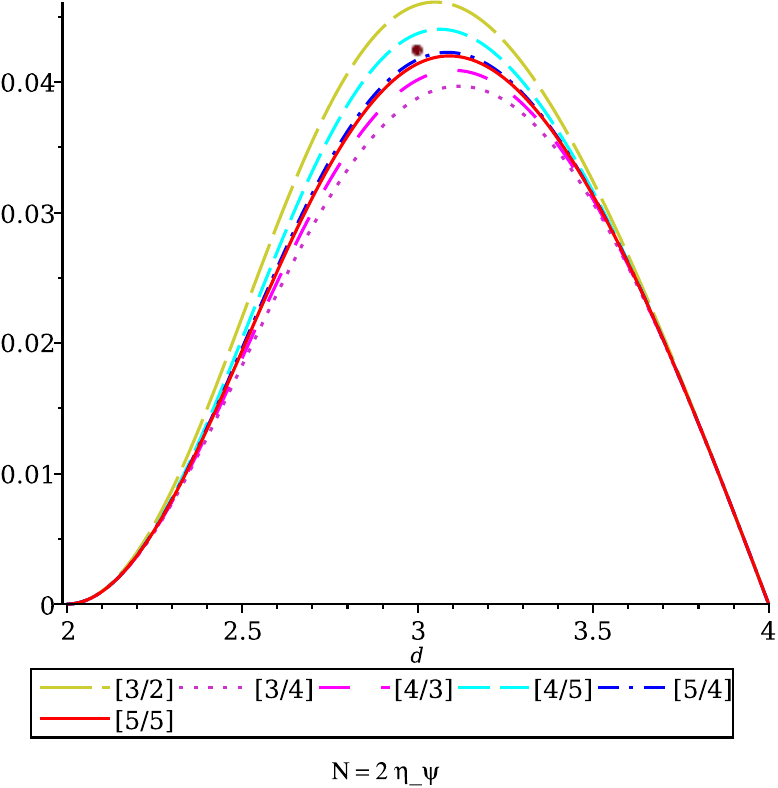}
\quad
\quad
\includegraphics[width=8.50cm,height=7.00cm]{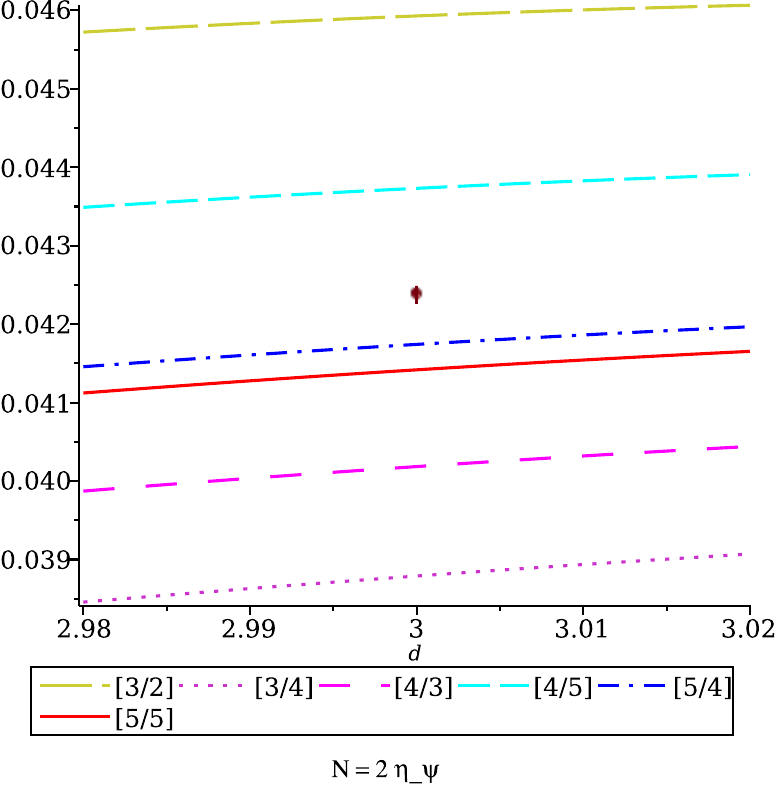}
\caption{Two-sided Pad\'{e} approximants for $\eta_\psi$ with $N$~$=$~$2$ in
$2$~$\leq$~$d$~$\leq$~$4$ (left panel) and $2.98$~$\leq$~$d$~$\leq$~$3.02$ 
(right panel)}
\label{etapsin8n2pad}
\end{figure}}

{\begin{figure}[hb]
\includegraphics[width=8.50cm,height=7.00cm]{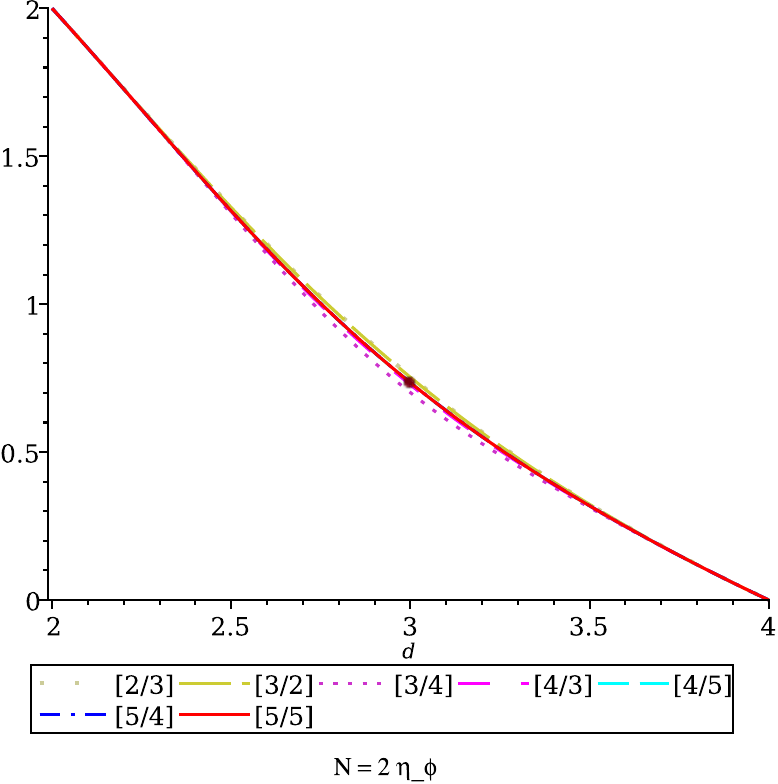}
\quad
\quad
\includegraphics[width=8.50cm,height=7.00cm]{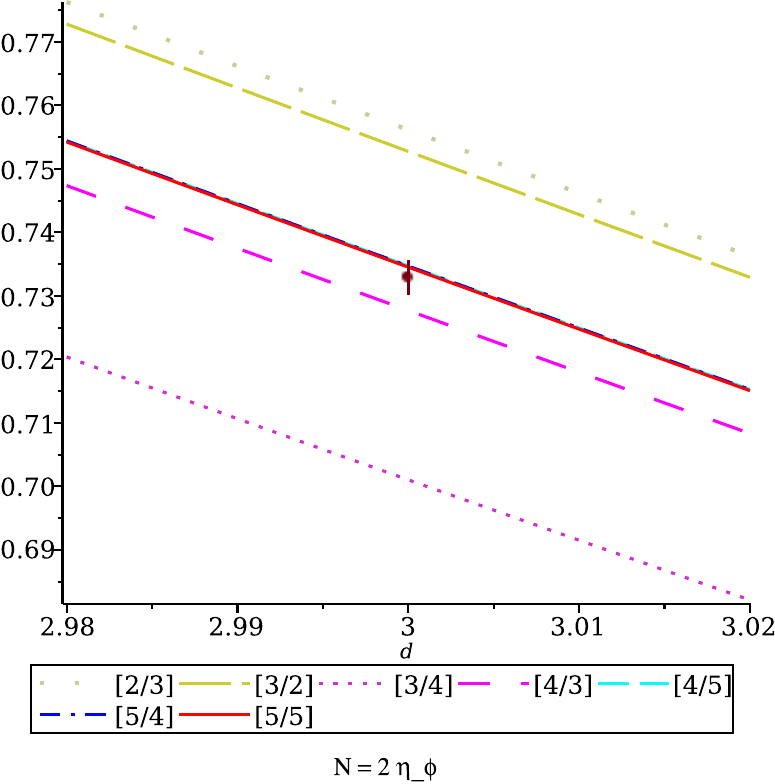}
\caption{Two-sided Pad\'{e} approximants for $\eta_\phi$ with $N$~$=$~$2$ in
$2$~$\leq$~$d$~$\leq$~$4$ (left panel) and $2.98$~$\leq$~$d$~$\leq$~$3.02$ 
(right panel)}
\label{etaphin8n2pad}
\end{figure}}

\subsection{Graphene case $N$~$=$~$2$}

Having summarized the two techniques employed to obtain exponent estimates we
now focus on three particular cases. The first concerns the CDW transition in
graphene corresponding to the value of $N$~$=$~$2$. Estimates for the two-sided
Pad\'{e} and interpolating polynomial are presented in Tables \ref{twopadn2} 
and \ref{intpolyn2} respectively. The two-sided Pad\'{e} approximants for 
$\eta_\phi$ were free from poles while those for $\eta_\psi$ were continuous 
for both approximations from three loops. For $1/\nu$ reliable estimates did 
not appear until four and five loops. In this case comparing with latest 
conformal bootstrap estimates recorded in the last two lines of Table 
\ref{sumn2} the $[5/5]$ value lies two $\sigma$ from the central value whereas 
that for $[5/4]$ is within one $\sigma$. Including two dimensional five loop 
information for $[5/6]$ and $[6,5]$ may mean they are accommodated within the 
bootstrap uncertainty. This is partly confirmed from the behaviour of 
$\eta_\phi$ as its estimates clearly progress with loop order to values within 
both bootstrap uncertainties. For $\eta_\psi$ its progression is not as settled
in comparison with the bootstrap. However we note that as the ultimate value is
generally accepted as small this means it is difficult to precisely measure 
which is partially reflected in the range of values in Table \ref{sumn2}. For 
the interpolating polynomial estimates of Table \ref{intpolyn2} we recall the 
$L$ loop estimates correspond to ${\cal I}_{[L,L]}$. While the four loops 
values were computed in \cite{Ihrig:2018hho} these were prior to 
\cite{Erramilli:2022kgp,Mitchell:2024hix}. So it is worth commenting on the 
estimates in relation to that more recent work. For both $1/\nu$ and 
$\eta_\phi$ their $[4,4]$ values are on the edge of both bootstrap 
uncertainties. Interestingly that for $\eta_\psi$ is also on the periphery of
the estimate of \cite{Erramilli:2022kgp}. We included ${\cal I}_{[L,L+1]}(3)$
estimates for each loop to gauge the effect of adding in five loop
information. For $[4,5]$ $1/\nu$ is on the upper edge of the bootstrap
uncertainty whereas $\eta_\phi$ is within that of \cite{Erramilli:2022kgp}. For
$\eta_\psi$ the situation is not as clear being $3\%$ different from the 
bootstrap value in contrast to the $0.5\%$ difference of the $[4,4]$ value.

{\begin{figure}[ht]
\includegraphics[width=8.50cm,height=7.00cm]{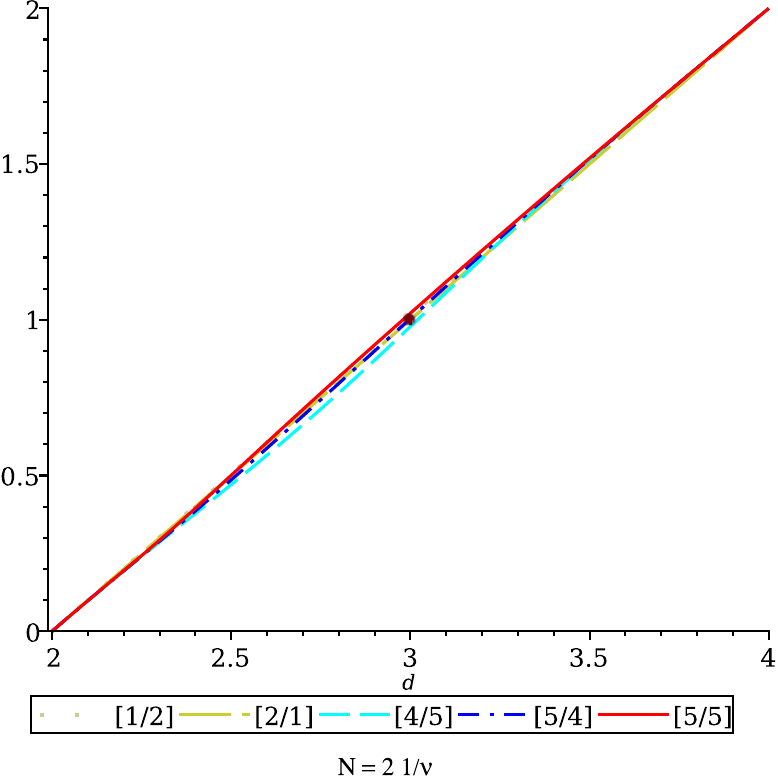}
\quad
\quad
\includegraphics[width=8.50cm,height=7.00cm]{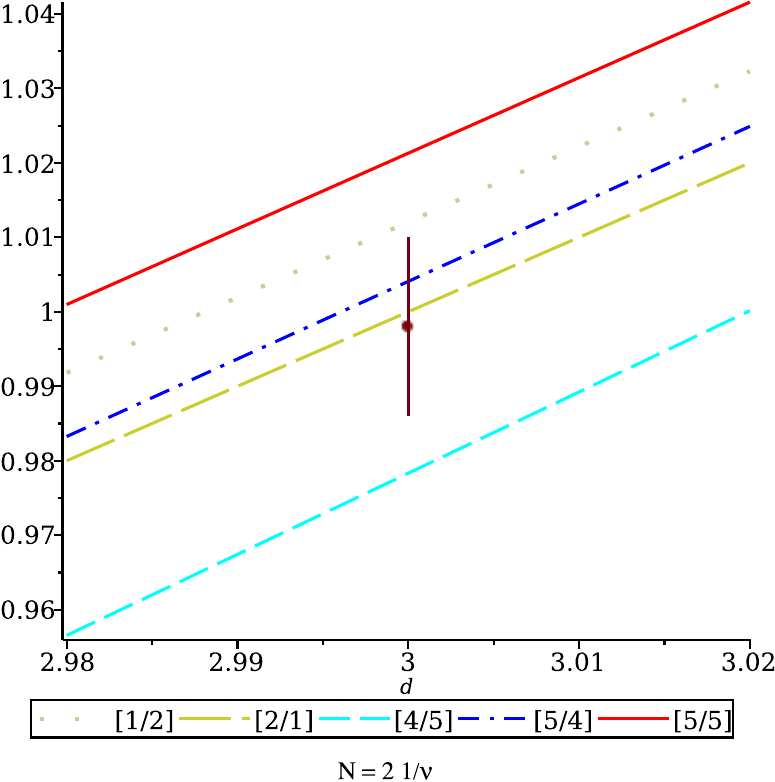}
\caption{Two-sided Pad\'{e} approximants for $1/\nu$ with $N$~$=$~$2$ in
$2$~$\leq$~$d$~$\leq$~$4$ (left panel) and $2.98$~$\leq$~$d$~$\leq$~$3.02$ 
(right panel)}
\label{nurecipn8n2pad}
\end{figure}}

{\begin{figure}[ht]
\includegraphics[width=8.50cm,height=7.00cm]{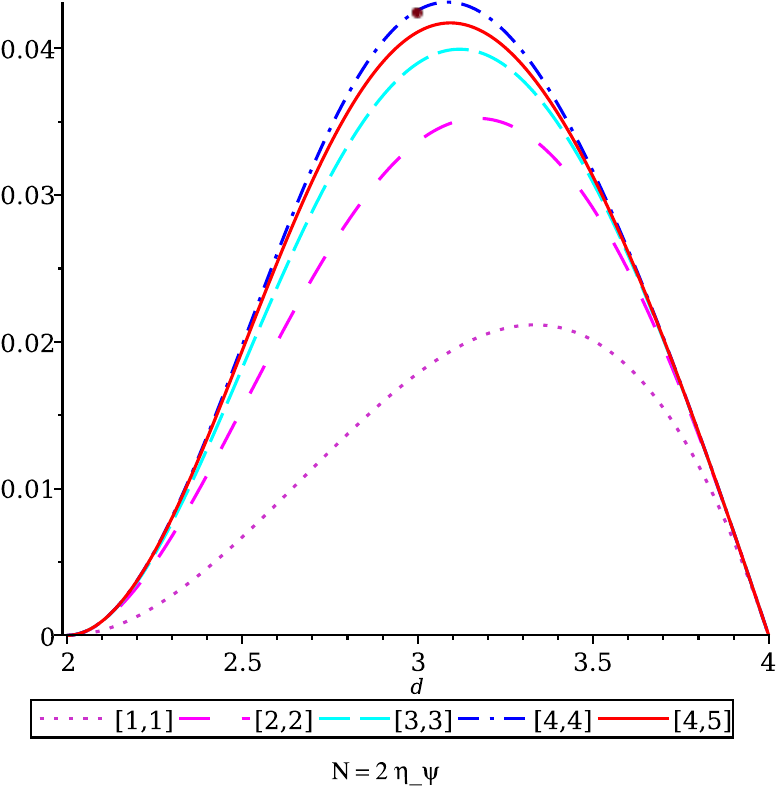}
\quad
\quad
\includegraphics[width=8.50cm,height=7.00cm]{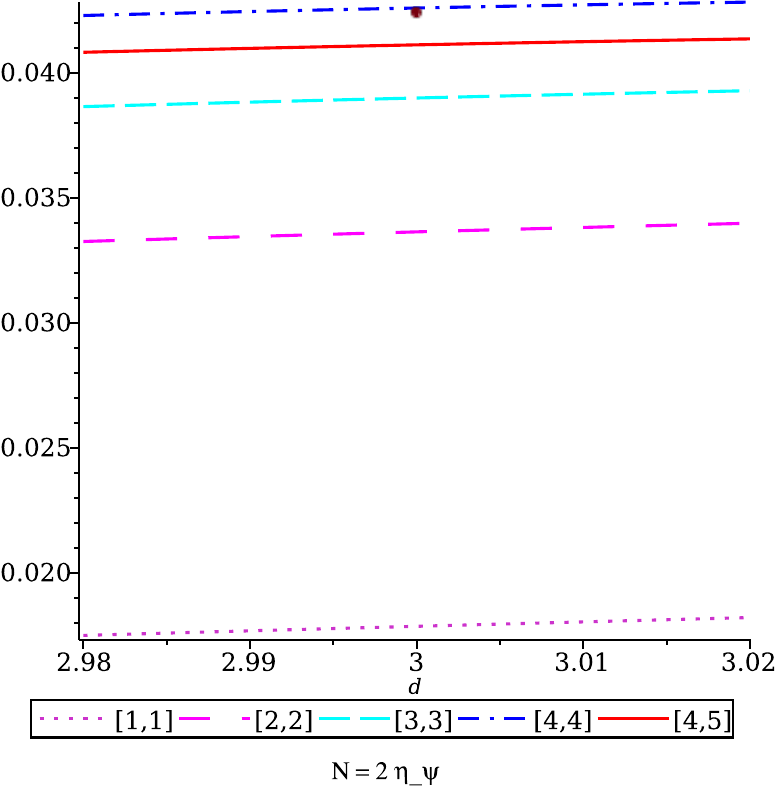}
\caption{Interpolating polynomials for $\eta_\psi$ with $N$~$=$~$2$ in
$2$~$\leq$~$d$~$\leq$~$4$ (left panel) and $2.98$~$\leq$~$d$~$\leq$~$3.02$ 
(right panel)}
\label{etapsin8n2pol}
\end{figure}}

While the tables indicate the numerical estimates with the increase of loop
order it is more instructive to appreciate the consequences graphically.
Therefore we have provided a set of plots for the two-sided Pad\'{e}
approximants in Figures \ref{etapsin8n2pad}, \ref{etaphin8n2pad} and
\ref{nurecipn8n2pad} with the analogous plots for the interpolating 
polynomial approach given in Figures \ref{etapsin8n2pol}, \ref{etaphin8n2pol}
and \ref{nurecipn8n2pol}. In each figure the plot on the left side represents
the $d$-dimensional estimate of the exponent in the full range 
$2$~$\leq$~$d$~$\leq$~$4$ whereas on each right hand side we focus on the
same plot but where $d$ lies between $2.98$ and $3.02$. This is because for
$\eta_\phi$ and $1/\nu$ the lines are virtually indistinguishable in the full
range. In addition we include a point in three dimensions which is the
conformal bootstrap estimate of \cite{Erramilli:2022kgp}. The error bars are
included for each of these points. However in the case of $\eta_\psi$ they are
relatively small compared to the actual point in the plots even in those 
focussed on the neighbourhood of three dimensions. For plots of the Pad\'{e}
approximants we included those for $[3/2]$ and higher loops for $\eta_\psi$
and $\eta_\phi$ but illustrated all five continuous and pole free ones for 
$1/\nu$. In the interpolating polynomial plots we present the behaviour of 
${\cal I}_{[L,L]}(d)$ as well as ${\cal I}_{[4,5]}(d)$. In all of the plots 
the solid line corresponds to the case where five loop information has been
included.

{\begin{figure}[ht]
\includegraphics[width=8.50cm,height=7.00cm]{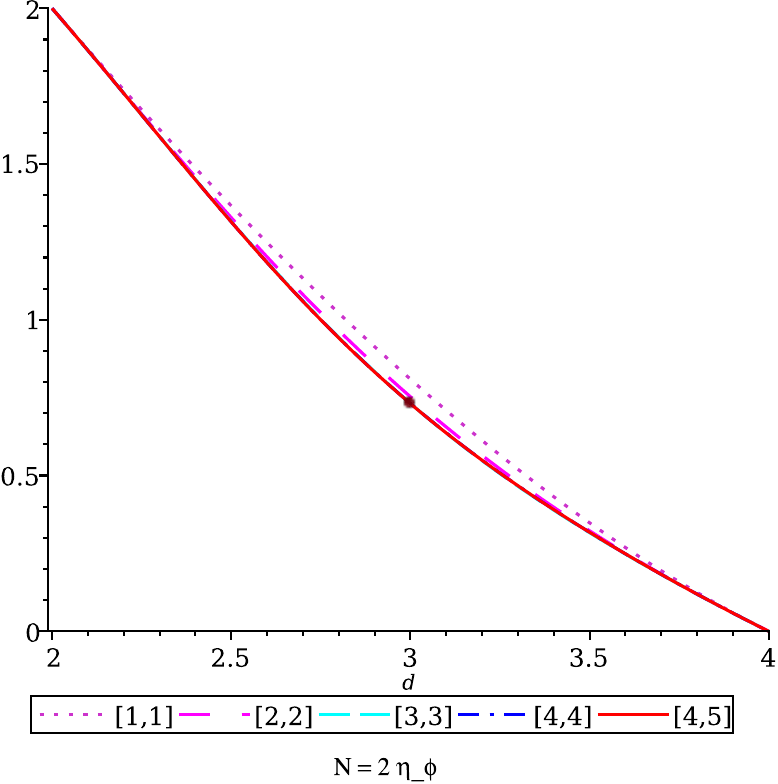}
\quad
\quad
\includegraphics[width=8.50cm,height=7.00cm]{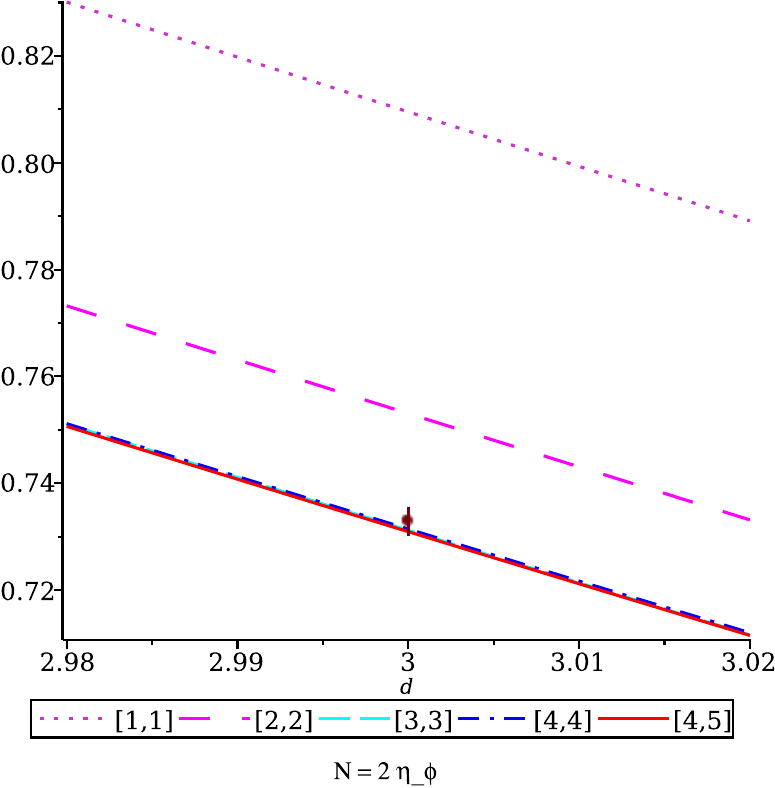}
\caption{Interpolating polynomials for $\eta_\phi$ with $N$~$=$~$2$ in
$2$~$\leq$~$d$~$\leq$~$4$ (left panel) and $2.98$~$\leq$~$d$~$\leq$~$3.02$ 
(right panel)}
\label{etaphin8n2pol}
\end{figure}}

{\begin{figure}[hb]
\includegraphics[width=8.50cm,height=7.00cm]{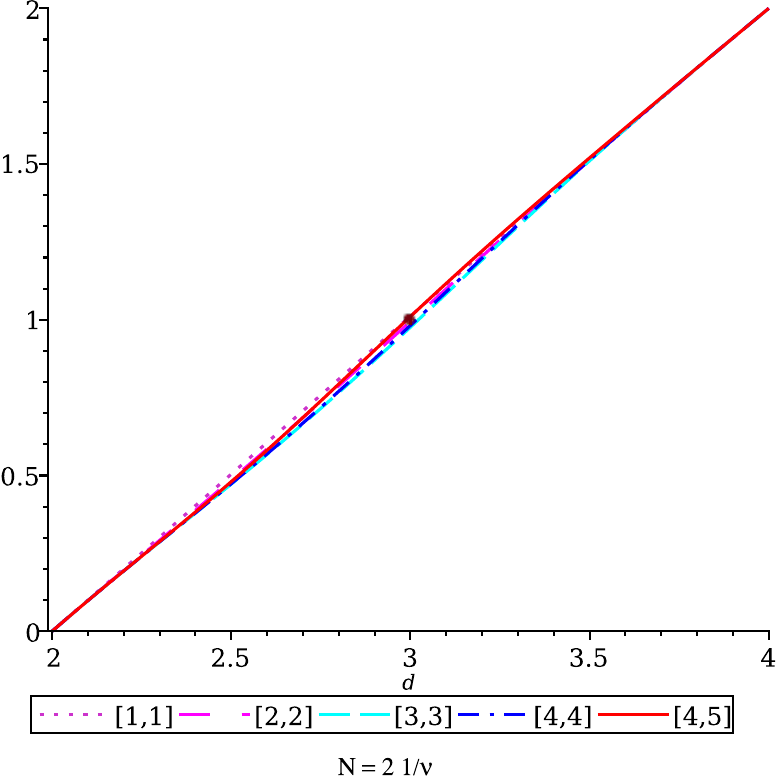}
\quad
\quad
\includegraphics[width=8.50cm,height=7.00cm]{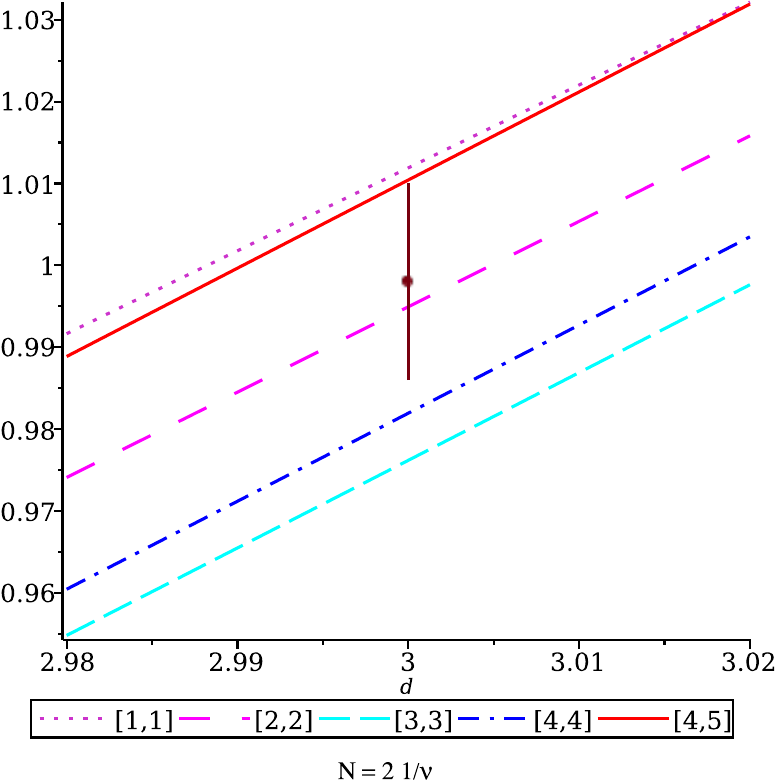}
\caption{Interpolating polynomials for $1/\nu$ with $N$~$=$~$2$ in
$2$~$\leq$~$d$~$\leq$~$4$ (left panel) and $2.98$~$\leq$~$d$~$\leq$~$3.02$ 
(right panel)}
\label{nurecipn8n2pol}
\end{figure}}

In general terms from the structure of the $d$-dimensional behaviour of the 
exponents using both techniques it is clear that as loop order increases there 
is at least qualitative convergence to expected properties. For instance $\psi$
is a free field in both the GN model and GNY theory meaning that $\eta_\psi$ in
$d$-dimensions has to have a turning point. In Figure \ref{etapsin8n2pad} the 
absence of the low loop approximants should not cloud the comparison with 
Figure \ref{etapsin8n2pol} which includes the one and two loop polynomials with
these corresponding to the lower two of the five lines. The higher loop plots 
in that case suggest there is oscillatory convergence towards the conformal 
bootstrap point which is more apparent in the focussed plot in Figure 
\ref{etapsin8n2pol}. Moreover there is no symmetry of the shape of the plots 
around three dimensions for either method as expected. For instance the 
tangents to $\eta_\psi$ at two and four dimensions have different slopes. In 
the neighbourhood of two dimensions $\eta_\psi$ is quadratic. Near four 
dimensions the gradient has to be negative for instance since $\eta_\psi$ 
extends beyond four dimensions towards the fermionic theory in six dimensions 
that is the ultraviolet extension of the GNY theory. For the theory in that
higher dimension $\psi$ will also be a free field. In principle if the 
renormalization group functions of its Lagrangian were available then 
additional information could be included in the Pad\'{e} and interpolating 
polynomial analyses. Although it is not clear if there would be a significant 
change in our three dimensional estimates. If one considers the gradient of the
tangent to $\eta_\psi$ in the region $3.5$~$\leq$~$d$~$\leq$~$3.9$ in Figures 
\ref{etapsin8n2pad} and \ref{etapsin8n2pol} and then extrapolate its trajectory
to estimate a value in three dimensions it is evident that it will exceed the 
exponent estimates. Repeating this exercise for the tangent in the range $2.3$ 
to $2.6$ then a similar overestimate will result. This in effect lies behind 
the overshoot of $\eta_\psi$ estimates given from the four dimensional
one-sided $[2/3]$ Pad\'{e} value in Table \ref{pade} as well as the two 
dimensional equivalent in Table \ref{sumn2}. The latter being larger than the
former overshoot is primarily driven by the tangent slopes in the regions 
either side of three dimensions. The behaviour of $\eta_\phi$ and $1/\nu$ 
differs significantly from $\eta_\psi$ as both are monotonic functions 
resulting from their two and four dimensional values being different integers. 
Indeed the $[2/1]$ Pad\'{e} approximant for $1/\nu$ is precisely the straight 
line $d$~$-$~$2$. The plots of each of these exponents from the two techniques 
show very similar behaviour in that the small differences in each approximation
only become evident in the plots around three dimensions. For $\eta_\phi$ the 
two highest loop order approximations touch the conformal bootstrap error bar 
and both lines are virtually on top of each other. While it is tempting to 
claim that there is good convergence that would have to wait until there is a 
five loop GN analysis. For $1/\nu$ the comparison between both techniques is 
not as concrete. If one ignores the low loop plots of ${\cal P}_{[1/2]}(d)$ and 
${\cal P}_{[2/1]}(d)$ in Figure \ref{nurecipn8n2pad} then the bootstrap 
point is bounded by ${\cal P}_{[4/5]}(d)$ and ${\cal P}_{[5/4]}(d)$. As 
${\cal P}_{[5/5]}(d)$ lies above ${\cal P}_{[5/4]}(d)$ any five loop input from
two dimensions to produce ${\cal P}_{[5/6]}(d)$ and ${\cal P}_{[6/5]}(d)$ would
have to lie within the four loop counterparts to point towards convergence near
the value of \cite{Erramilli:2022kgp}. By contrast in Figure 
\ref{nurecipn8n2pol} ${\cal I}_{[3,3]}(d)$ and ${\cal I}_{[4,4]}(d)$ appear to 
be creeping towards the lower edge of the conformal bootstrap estimate but the 
${\cal I}_{[4,5]}(d)$ curve overshoots. Again a better understanding of the 
trend would require a full five loop comparison.

{\begin{table}[ht]
\begin{center}
\begin{tabular}{|c||c|c|c|}
\hline
Method and source & $\eta_\psi$ & $\eta_\phi$ & $1/\nu$ \\
\hline
Large $N$ \cite{Gracey:1992cp,Vasiliev:1993pi,Vasiliev:1992wr,Gracey:1993kb,
Gracey:1993kc} & $0.1056$ & $0.509$ & $0.961$ \\
Four loop two-sided Pad\'{e} \cite{Ihrig:2018hho} & $(0.086)$ & $(0.480)$ & $(0.961)$ \\
$~$Four loop interpolating polynomial \cite{Ihrig:2018hho}$~$ & $0.140$ & 
$0.397$ & $1.040$ \\
Four loop Pad\'{e}-Borel \cite{Ihrig:2018hho} & $0.102(12)$ & $0.487(12)$ & $1.114(33)$ \\
Monte Carlo \cite{Li:2014aoa} & ------ & $0.45(2)$ & $1.30(5)$ \\
Functional renormalization group \cite{Knorr:2016sfs} & $0.0654$ & $0.5506$ & 
$1.075(4)$ \\
Three loop $d$~$=$~$4$ naive Pad\'{e} \cite{Mihaila:2017ble} & $0.102$ & 
$0.463$ & $1.166$ \\
Conformal bootstrap \cite{Iliesiu:2017nrv} & $0.084$ & $0.544$ & $0.76$ \\
Monte Carlo \cite{Huffman:2017swn} & ------ & $0.54(6)$ & $1.14(2)$ \\
Four loop $d$~$=$~$4$ naive Pad\'{e} \cite{Zerf:2017zqi} & $0.0976$ & 
$0.4969$ & ------ \\
Four loop $d$~$=$~$4$ naive Pad\'{e} \cite{Zerf:2017zqi} & $0.0972$ & 
$0.4872$ & ------ \\
Four loop $d$~$=$~$4$ naive Pad\'{e} \cite{Zerf:2017zqi} & ------ & ------ & 
$1.101$ \\
Monte Carlo \cite{Huffman:2019efk} & ------ & $0.51(3)$ & $1.124(13)$ \\
Conformal bootstrap \cite{Erramilli:2022kgp} & $~0.08712(32)~$ & 
$~0.5155(30)~$ & $~1.101(10)~$ \\
Conformal bootstrap \cite{Mitchell:2024hix} & ------ & $0.5156(30)$ & 
$1.101(10)$ \\
\hline
\end{tabular}
\caption{Summary of previous exponent estimates for $N$~$=$~$1$.}
\label{sumn1}
\end{center}
\end{table}}

\subsection{Other cases}

Although the $N$~$=$~$2$ GNY theory has a major physical application several 
other values of $N$ are of interest and we briefly summarize a similar analysis
to that of graphene for $N$~$=$~$1$ and $5$. The former governs semi-metal
transitions for spinless fermions on a honeycomb lattice whereas we consider
the latter case in order to compare with conformal bootstrap and functional
renormalization group results. First for $N$~$=$~$1$ we provide a summary of
exponent estimates from a variety of techniques in Table \ref{sumn1}. It has
parallels to the $N$~$=$~$2$ summary table in that estimates for $\eta_\psi$
have a wide range whereas there is mostly a better consensus for $\eta_\phi$ 
and $1/\nu$. Tables \ref{twopadn1} and \ref{intpolyn1} record our two-sided
Pad\'{e} and interpolating polynomial estimates. The former is more sparse
than the graphene case which is due to poles appearing in 
$2$~$\leq$~$d$~$\leq$~$4$ with only one such instance occurring for 
$\eta_\phi$. The incompleteness of estimates for all three exponents prevents 
any general convergence trends emerging. Instead all we can deduce is that 
there is a broad general agreement with results provided in Table \ref{sumn1}. 
Although the interpolating polynomial method has by construction no continuity 
problems the $N$~$=$~$1$ estimates also do not show a convergence path. Only 
the higher loop estimates for $1/\nu$ are within previous results of Table 
\ref{sumn1}. For $\eta_\phi$ the ${\cal I}_{[i,j]}(3)$ values are 
unsurprisingly more in line with other perturbative results rather than 
conformal bootstrap, Monte Carlo and functional renormalization group 
techniques. It may be the case that several more orders in perturbation theory 
would be required to obtain commensurate estimates.

{\begin{table}[ht]
\begin{minipage}{7.0cm}
\centering
\begin{tabular}{|c||c|c|c|}
\hline
Pad\'{e} & $\eta_\psi$ & $\eta_\phi$ & $1/\nu$ \\
\hline
$~[1/2]~$ & ------ & $~0.615385~$ & $~1.043101~$ \\
$[2/1]$ & ------ & $0.625000$ & $1.000000$ \\
$[2/3]$ & ------ & $0.564955$ & ------ \\
$[3/2]$ & $0.107695$ & $0.562831$ & ------ \\
$[3/4]$ & $0.074744$ & $0.401109$ & $0.922860$ \\
$[4/3]$ & $0.090275$ & $0.479073$ & ------ \\
$[4/5]$ & $0.098936$ & $0.479788$ & ------ \\
$[5/4]$ & ------ & ------ & $0.961217$ \\
$[5/5]$ & ------ & $0.507009$ & $1.120413$ \\
\hline
\end{tabular}
\caption{Two-sided Pad\'{e} exponent estimates ${\cal P}_{[m/n]}(3)$ for 
$N$~$=$~$1$.}
\label{twopadn1}
\end{minipage}
\hspace{2.0cm}
\begin{minipage}{7.0cm}
\centering
\begin{tabular}{|c||c|c|c|}
\hline
Polynomial & $\eta_\psi$ & $\eta_\phi$ & $1/\nu$ \\
\hline
$~[1,1]~$ & $~0.025000~$ & $~0.600000~$ & $~1.041320~$ \\
$[1,2]$ & $0.052560$ & $0.600613$ & $1.081212$ \\
$[2,2]$ & $0.079405$ & $0.419057$ & $0.988436$ \\
$[2,3]$ & $0.072515$ & $0.489743$ & $1.042067$ \\
$[3,3]$ & $0.116265$ & $0.371511$ & $0.950413$ \\
$[3,4]$ & $0.097376$ & $0.450625$ & $1.005131$ \\
$[4,4]$ & $0.139895$ & $0.397030$ & $1.039686$ \\
$[4,5]$ & $0.111856$ & $0.435192$ & $1.072936$ \\
\hline
\end{tabular}
\caption{Interpolating polynomial ${\cal I}_{[i,j]}(3)$ exponent estimates for 
$N$~$=$~$1$.}
\label{intpolyn1}
\end{minipage}
\end{table}}

Turning to the case of $N$~$=$~$5$ there has not been the same volume of
activity to extract critical exponents both analytically and numerically. This
is reflected in the summary of the current situation in Table \ref{sumn5}. 
Unlike the $N$~$=$~$1$ case there is a greater general consensus as to the
rough values of the three exponents. However all estimates employ continuum
field theory approaches as there are no results from lattice field theory
or discrete models which would produce Monte Carlo values. The results of our
two-sided Pad\'{e} and interpolating polynomial methods are displayed in
Tables \ref{twopadn5} and \ref{intpolyn5} respectively. For the former more
approximants are continuous and pole free compared to $N$~$=$~$1$ with a full
set available for $\eta_\phi$. From examining the singular Pad\'{e}
approximants for each of our three values of $N$ it is generally the case that 
either a simple pole appears in $2$~$<$~$d$~$<$~$4$ above two dimensions or 
there is a pair of poles in the regions either side of three dimensions. Then
by looking at the $d$ dependence for the same approximants as $N$ increases the
poles above two dimensions move towards the origin or evaporate. Equally where 
there is a simple pole above three dimensions it translates towards six 
dimensions. We note that the absence of $[3/4]$ and $[5/4]$ approximants for 
$1/\nu$ when $N$~$=$~$5$ is due to the presence of a zero and simple pole which
are located within $10^{-4}$ or less of each other. This was an unusual 
occurrence across all our Pad\'{e} analyses. For the interpolating polynomial 
estimates there were obviously no such hindrances and values listed in Table 
\ref{intpolyn5} show a converging trend for $\eta_\psi$ and $\eta_\phi$. For 
$1/\nu$ there is a similar direction except that the final value increases 
indicating that a lower dimensional computation would be needed to clarify the 
situation. Both techniques however show good consistency with the various 
estimates given in Table \ref{sumn5} except for the $\eta_\psi$ functional
renormalization group value from \cite{Ihrig:2018hho}.

{\begin{table}[ht]
\begin{center}
\begin{tabular}{|c||c|c|c|}
\hline
Method and source & $\eta_\psi$ & $\eta_\phi$ & $1/\nu$ \\
\hline
Large $N$ \cite{Gracey:1992cp,Vasiliev:1993pi,Vasiliev:1992wr,Gracey:1993kb,
Gracey:1993kc} & $0.0152$ & $0.894$ & $0.970$ \\
Conformal bootstrap \cite{Iliesiu:2017nrv} & $0.016$ & $0.888$ & $0.97$ \\
Four loop two-sided Pad\'{e} \cite{Ihrig:2018hho} & $0.0151$ & $0.893$ & 
$0.9840$ \\
$~$Four loop interpolating polynomial \cite{Ihrig:2018hho}$~$ & $0.0150$ & 
$0.893$ & $0.9763$ \\
Four loop Pad\'{e}-Borel \cite{Ihrig:2018hho} & $~0.0120(48)~$ & 
$~0.893(9)~$ & $~0.9580(75)~$ \\
Functional renormalization group \cite{Ihrig:2018hho} & $0.011$ &
$0.911$ & $0.980$ \\
\hline
\end{tabular}
\caption{Summary of previous exponent estimates for $N$~$=$~$5$.}
\label{sumn5}
\end{center}
\end{table}}

{\begin{table}[ht]
\begin{minipage}{7.0cm}
\centering
\begin{tabular}{|c||c|c|c|}
\hline
Pad\'{e} & $\eta_\psi$ & $\eta_\phi$ & $1/\nu$ \\
\hline
$~[1/2]~$ & ------ & $~0.906801~$ & $~0.991323~$ \\
$[2/1]$ & ------ & $0.925000$ & $1.000000$ \\
$[2/3]$ & ------ & $0.899023$ & $0.990855$ \\
$[3/2]$ & $0.016010$ & $0.895482$ & $0.989372$ \\
$[3/4]$ & $0.014678$ & $0.889395$ & ------ \\
$[4/3]$ & $0.014680$ & $0.891326$ & $0.980097$ \\
$[4/5]$ & $0.015218$ & $0.892547$ & $0.976555$ \\
$[5/4]$ & $0.015055$ & $0.892462$ & ------ \\
$[5/5]$ & $0.015070$ & $0.892288$ & $0.979693$ \\
\hline
\end{tabular}
\caption{Two-sided Pad\'{e} exponent estimates ${\cal P}_{[m/n]}(3)$ for 
$N$~$=$~$5$.}
\label{twopadn5}
\end{minipage}
\hspace{2.0cm}
\begin{minipage}{7.0cm}
\centering
\begin{tabular}{|c||c|c|c|}
\hline
Polynomial & $\eta_\psi$ & $\eta_\phi$ & $1/\nu$ \\
\hline
$~[1,1]~$ & $~0.009615~$ & $~0.914530~$ & $~0.991247~$ \\
$[1,2]$ & $0.015642$ & $0.896196$ & $0.991156$ \\
$[2,2]$ & $0.013890$ & $0.898567$ & $0.986445$ \\
$[2,3]$ & $0.015018$ & $0.892908$ & $0.980349$ \\
$[3,3]$ & $0.014535$ & $0.893099$ & $0.977654$ \\
$[3,4]$ & $0.015041$ & $0.893362$ & $0.977896$ \\
$[4,4]$ & $0.015013$ & $0.892816$ & $0.976292$ \\
$[4,5]$ & $0.015041$ & $0.892008$ & $0.980487$ \\
\hline
\end{tabular}
\caption{Interpolating polynomial ${\cal I}_{[i,j]}(3)$ exponent estimates for 
$N$~$=$~$5$.}
\label{intpolyn5}
\end{minipage}
\end{table}}

\section{Conclusions}
\label{sectconc}

One of the main aims of our study was to establish the five loop 
renormalization group functions of the four dimensional GNY model with a 
multiplet of $N$ fermions. In achieving this level of accuracy we have been
able to apply the results to improve estimates of critical exponents for phase 
transitions in materials such as graphene by using several summation 
approaches. These exploit the $d$-dimensional connection of the GN and GNY 
models that guide the $\epsilon$ expansion along a reliable path to three 
dimensions. 

Previous lower loop results, \cite{Mihaila:2017ble, Zerf:2017zqi, 
Ihrig:2018hho}, had shown an encouraging trend towards agreement with other 
methods such as the functional renormalization group technology and the 
conformal bootstrap approach. In the last few years improved bootstrap 
exponents have become available, \cite{Erramilli:2022kgp, Mitchell:2024hix}, so
that it was necessary to move the perturbative analysis to the next loop order.
The first phase of this has now been completed here with updated two-sided 
Pad\'{e} approximant and interpolating polynomial constructions in 
$d$-dimensions. The second phase would require five loop GN renormalization 
group functions to fully refine the new dimension four based analysis carried 
out here. 

In respect of that the general picture for $N$~$=$~$2$ appears to be similar to
the conclusions of the four loop analysis of \cite{Ihrig:2018hho} in that there
is reasonable agreement with the latest bootstrap exponent estimates. For 
instance for $\eta_\phi$ and to a slightly lesser degree $1/\nu$ the estimates 
are on the edge of the uncertainty bands of \cite{Erramilli:2022kgp, 
Mitchell:2024hix}. The situation with $\eta_\psi$ is harder to pin down given 
that its value is small being of the magnitude of a few hundredths. Despite 
this our two perturbative summations, especially the interpolating polynomial 
one, are edging closer to the latest $\eta_\psi$ conformal bootstrap value. 
This is an encouraging motivation to complete the five loop analysis of the 
universality class started here by renormalizing the GN model to the same 
order.

\begin{acknowledgments}
The work of J.A.G. was carried out with the support of the STFC Consolidated 
Grant ST/X000699/1 and a Visiting Scholarship from the Kolleg Mathematik Physik 
Berlin.
The work of A.M. was supported in part by the Spanish Ministry of Science and
Innovation (PID2020-112965GB-I00,PID2023-146142NB-I00), and by the Departament 
de Recerca i Universities from Generalitat de Catalunya to the Grup de Recerca 
00649 (Codi: 2021 SGR 00649). This project received funding from the European 
Union’s Horizon 2020 research and innovation programme under grant agreement No
824093. IFAE is partially funded by the CERCA program of the Generalitat de 
Catalunya.
Y.S. acknowledges support from ANID under FONDECYT project No. 1231056 and from
UBB/VRIP project No. EQ2351247.
For the purpose of open access, the authors have
applied a Creative Commons Attribution (CC-BY) licence to any Author Accepted
Manuscript version arising.
\end{acknowledgments}

\appendix

\section{Large $N$ critical exponents}
\label{applargen}

In order to carry out the large $N$ check of the exponents by comparing with
those derived from the five loop results, we record the explicit expressions 
for the relevant large $N$ critical exponents. These have been determined in 
$d$~$=$~$2\mu$ dimensions in a variety of articles. For instance the fermion 
anomalous dimension is available to $\order(1/N^3)$ \cite{Gracey:1990wi,
Vasiliev:1992wr, Gracey:1993kc} being given by
\begin{eqnarray}
\eta_\psi &=&
\frac{\eta_1}{N}
+ \left( \frac{1}{2 \mu} + \frac{1}{(\mu-1)^2} + \frac{1}{2 (\mu-1)}
+ \frac{3}{(\mu-2)} + \frac{4}{(2 \mu-3)}
+ \frac{(2 \mu-1)}{(\mu-1)} \hat{\Psi}(\mu) \right) \frac{\eta_1^2}{N^2}
\nonumber \\
&&
+ \left(
\frac{1}{2 \mu^2} - \frac{9}{2} - \frac{7}{2 \mu} - \frac{5}{4 (\mu-1)^4}
- \frac{3}{(\mu-1)^3} - \frac{51}{4 (\mu-1)^2} - \frac{79}{4 (\mu-1)}
+ \frac{9}{(\mu-2)^2} + \frac{153}{4 (\mu-2)} + \frac{16}{(2 \mu-3)^2}
\right. \nonumber \\
&& \left. ~~~~
-~ \frac{52}{(2 \mu-3)} - \mu
+ \left( 6 + \frac{5}{4 (\mu-1)^2} - \frac{3}{(\mu-1)} + \frac{12}{(\mu-2)}
+ \frac{27}{2 (2 \mu-3)} + \frac{1}{4} \mu \right) \hat{\Theta}(\mu)
+ \frac{3 \mu^2}{(\mu-1)} \hat{\Psi}(\mu) \hat{\Theta}(\mu)
\right. \nonumber \\
&& \left. ~~~~
+~ \left( \frac{3}{2 \mu} - \frac{3}{2} + \frac{1}{(\mu-1)^3}
+ \frac{2}{(\mu-1)^2} - \frac{22}{(\mu-1)} + \frac{27}{(\mu-2)}
+ \frac{48}{(2 \mu-3)} - \mu \right) \hat{\Psi}(\mu)
+ \frac{3 \mu^2}{2 (\mu-1)} \Xi(\mu) \hat{\Theta}(\mu)
\right. \nonumber \\
&& \left. ~~~~
+~ \frac{(2 \mu-1)^2}{2 (\mu-1)^2} \left( \hat{\Phi}(\mu) + 3 \hat{\Psi}^2(\mu)
\right) \right) \frac{\eta_1^3}{N^3} ~+~ \order \left( \frac{1}{N^4} \right) 
\end{eqnarray}
where \cite{Hands:1992be,Gracey:1990wi}
\begin{equation}
\eta_1 ~=~ -~ \frac{\Gamma(2 \mu-1)}{2 \mu \Gamma^2(\mu) \Gamma(\mu-1)
\Gamma(1-\mu)} ~.
\end{equation}
The expressions for $\eta_\psi$ and the remaining exponents are recorded with
the four dimensional spinor trace convention for consistency although several 
of the original results were constructed in relation to the two dimensional GN 
model convention. We have introduced the shorthand notation
\begin{eqnarray}
\hat{\Psi}(\mu) &=& \psi(2 \mu-3) ~-~ \psi(1) ~+~ \psi(3-\mu) ~-~ \psi(\mu-1)
\nonumber \\
\hat{\Phi}(\mu) &=& \psi^\prime(2 \mu-3) ~-~ \psi^\prime(3-\mu) ~-~
\psi^\prime(\mu-1) ~+~ \psi^\prime(1)
\nonumber \\
\hat{\Theta}(\mu) &=& \psi^\prime(\mu-1) ~-~ \psi^\prime(1)
\end{eqnarray}
where the $\epsilon$ expansion of each function near four dimensions involves
$\zeta_n$ only. At $\order(1/N^3)$ the function $\Xi(\mu)$ appears which is 
related to a specific two loop Feynman integral introduced in 
\cite{Vasiliev:1982dc} and given in equation (4.13) of \cite{Gracey:1993kc} in 
relation to the GN universality class. When $\mu$~$=$~$2$~$-$~$\half \epsilon$ 
then the combination $\Xi(\mu) \hat{\Theta}(\mu)$ has the expansion
\begin{eqnarray}
\Xi(\mu) \hat{\Theta}(\mu) &=&
\frac{2}{3} \zeta_3 \epsilon
+ \left(
\frac{1}{3} \zeta_3
+ \frac{1}{2} \zeta_4
- \frac{5}{4} \zeta_5
\right) \epsilon^2
+ \left(
\frac{1}{3} \zeta_5
- \frac{25}{16} \zeta_6
+ \frac{1}{4} \zeta_4
+ \frac{1}{6} \zeta_3
+ \frac{5}{8} \zeta_3^2
\right) \epsilon^3
\nonumber \\
&&
+ \left(
\frac{5}{24} \zeta_6
- \frac{121}{64} \zeta_7
+ \frac{1}{6} \zeta_5
+ \frac{1}{8} \zeta_4
+ \frac{1}{12} \zeta_3
+ \frac{15}{16} \zeta_3 \zeta_4
\right) \epsilon^4
\nonumber \\
&&
+ \left(
\frac{9}{80} \zeta_{5,3}
- \frac{577}{320} \zeta_8
+ \frac{1}{8} \zeta_7
+ \frac{5}{48} \zeta_6
+ \frac{1}{12} \zeta_5
+ \frac{1}{16} \zeta_4
+ \frac{1}{24} \zeta_3
+ \frac{9}{8} \zeta_3 \zeta_5
\right) \epsilon^5 ~+~ \order(\epsilon^6)
\label{xithet}
\end{eqnarray}
which was deduced from the discussion given in \cite{Broadhurst:1996ur,
Broadhurst:1996yc} and $\zeta_{5,3}$ is a multiple zeta. The critical exponent 
relating to the scalar field is, \cite{Vasiliev:1993pi, Gracey:1993kb},
\begin{eqnarray}
\eta_\phi &=& 4 ~-~ 2 \mu
-~ \frac{2 (2 \mu-1)}{(\mu-1)} \frac{\eta_1}{N}
\nonumber \\
&& + \left( 6 - \frac{1}{\mu} + \frac{5}{(\mu-1)^3} + \frac{13}{(\mu-1)^2}
+ \frac{29}{(\mu-1)} - \frac{18}{(\mu-2)} - \frac{32}{(2 \mu-3)} + 4 \mu
- \frac{6 \mu^2}{(\mu-1)} \hat{\Theta}(\mu)
- \frac{2 (2 \mu-1)^2}{(\mu-1)^2} \hat{\Psi}(\mu) \right) 
\frac{\eta_1^2}{N^2}
\nonumber \\
&& +~ \order \left( \frac{1}{N^3} \right)
\end{eqnarray}
at $\order(1/N^2)$ while the dimension of its mass operator is, 
\cite{Vasiliev:1993pi,Gracey:1993kb},
\begin{eqnarray}
\eta_{\phi^2} &=&
-~ \frac{2 \mu (2 \mu-1)}{(\mu-1)} \frac{\eta_1}{N}
\nonumber \\
&&
+ \left(  \frac{ 2 \mu}{(\mu-1) (\mu-2)^2 \eta_1}
- 38 + \frac{38}{(\mu-1)^2} + \frac{134}{(\mu-1)} + \frac{8}{(\mu-2)^3} 
- \frac{16}{(\mu-2)^2} - \frac{182}{(\mu-2)}
- \frac{144}{(2 \mu-3)} - 16 \mu + 8 \mu^2
\right. \nonumber \\
&& \left. ~~~~
-~ \frac{3 \mu^2 (2 \mu^2-11 \mu+8)}{(\mu-1) (\mu-2)} \hat{\Theta}(\mu)
- \left( 38 + \frac{10}{(\mu-1)^2} + \frac{14}{(\mu-1)}
+ \frac{24}{(\mu-2)^2} + \frac{80}{(\mu-2)} + 8 \mu + 8 \mu^2 \right)
\hat{\Psi}(\mu)
\right. \nonumber \\
&& \left. ~~~~
-~ \frac{4 \mu^2 (2 \mu-3)}{(\mu-1) (\mu-2)} 
\left( \hat{\Phi}(\mu) + \hat{\Psi}^2(\mu) \right) \right) 
\frac{\eta_1^2}{N^2} ~+~ \order \left( \frac{1}{N^3} \right) ~.
\end{eqnarray}
The remaining two exponents, $\omega_\pm$, correspond to the eigen-anomalous
dimensions of the matrix of critical $\beta$-function slopes $\beta_{ij}$
discussed earlier. These were computed at $\order(1/N^2)$ in the large $N$
expansion in \cite{Manashov:2017rrx}, where the subtlety of mixing at the large
$N$ critical point was discussed, extending the leading order expression for 
$\omega_-$ given in \cite{Gracey:2017fzu}. Although the $d$-dimensional 
$\order(1/N^2)$ exponents were given implicitly in \cite{Manashov:2017rrx} we 
record the full expressions here partly for future reference but also to ensure
our trace conventions are consistent in checking large $N$ results with the 
perturbative $\epsilon$ expansion for all exponents. We have
\begin{eqnarray}
\omega_+ &=&
4 - 2 \mu + \frac{4 (2 \mu-1) (3 \mu-1)}{(\mu-1)} \frac{\eta_1}{N}
\nonumber \\
&&
+ \left( 
24 - \frac{2}{\mu} - \frac{38}{(\mu-1)^3} - \frac{394}{(\mu-1)^2}
- \frac{1106}{(\mu-1)} - \frac{48}{(\mu-2)^3} + \frac{96}{(\mu-2)^2}
+ \frac{1056}{(\mu-2)} + \frac{800}{(2 \mu-3)} - 16 \mu - 192 \mu^2
\right. \nonumber \\
&& \left. ~~~~
-~ \frac{12 \mu}{(\mu-1) (\mu-2)^2 \eta_1}
+ \left( 212 + \frac{56}{(\mu-1)^2} + \frac{68}{(\mu-1)}
+ \frac{144}{(\mu-2)^2} + \frac{480}{(\mu-2)} + 48 \mu + 48 \mu^2 \right)
\hat{\Psi}(\mu)
\right. \nonumber \\
&& \left. ~~~~
+~ \frac{54 \mu^2 (2 \mu^2-7 \mu+4)}{(\mu-1) (\mu-2)} \hat{\Theta}(\mu)
+ \frac{24 \mu^2 (2 \mu-3)}{(\mu-1) (\mu-2)}
\left( \hat{\Phi}(\mu) + \hat{\Psi}^2(\mu) \right) \right)
\frac{\eta_1^2}{N^2} ~+~ \order \left( \frac{1}{N^3} \right)
\end{eqnarray}
and
\begin{eqnarray}
\omega_- &=&
4 - 2 \mu
+ \frac{4 (2 \mu-1) (\mu-2)}{(\mu-1)} \frac{\eta_1}{N}
\nonumber \\
&&
+ \left(
88 - \frac{19}{\mu} + \frac{9}{(\mu-1)^3} + \frac{1}{(\mu-1)^2}
+ \frac{39}{(\mu-1)} + \frac{20}{(\mu-2)^2} + \frac{54}{(\mu-2)}
- \frac{18}{(\mu-3)} - \frac{32}{(2 \mu-3)} + 30 \mu - 32 \mu^2 + 8 \mu^3
\right. \nonumber \\
&& \left. ~~~~
-~ \frac{2 (2\mu-3) (2\mu^3-6 \mu^2+11 \mu-9)}{(\mu-1) (\mu-2) (\mu-3) \eta_1}
+ \frac{3 \mu (4 \mu^3-14 \mu^2+16 \mu-9)}{(\mu-1)} \hat{\Theta}(\mu)
\right. \nonumber \\
&& \left. ~~~~
+~ \left( 24 - \frac{14}{(\mu-1)} + \frac{20}{(\mu-2)} + \frac{36}{(\mu-3)}
- 24 \mu + 32 \mu^2 \right) \hat{\Psi}(\mu) \right) \frac{\eta_1^2}{N^2} ~+~ 
\order \left( \frac{1}{N^3} \right) ~.
\end{eqnarray}
If we expand each exponent in powers of $\epsilon$ near four dimensions and
compare with the corresponding exponents from the explicit five loop 
computation we find total agreement to the above respective orders in $1/N$.
Essential for the $\eta_\psi$ check at four and five loops at $\order(1/N^3)$ 
were the $\order(\epsilon^3)$ and higher terms of $\Xi(\mu) \hat{\Theta}(\mu)$ 
in (\ref{xithet}). For completeness and for future higher order computations we
note the $\epsilon$ expansion for $\eta_\psi$ is
\begin{eqnarray}
\eta_\psi &=&
\left(
\frac{1}{4} \epsilon
- \frac{3}{16} \epsilon^2
- \frac{3}{64} \epsilon^3
+ \left(
\frac{1}{16} \zeta_3
- \frac{3}{256}
\right) \epsilon^4
+ \left(
\frac{3}{64} \zeta_4
- \frac{3}{64} \zeta_3
- \frac{3}{1024}
\right) \epsilon^5
+ \left(
\frac{3}{64} \zeta_5
- \frac{9}{256} \zeta_4
- \frac{3}{256} \zeta_3
- \frac{3}{4096}
\right) \epsilon^6
\right. \nonumber \\
&& \left. ~
+ \left(
\frac{1}{128} \zeta_3^2
+ \frac{5}{128} \zeta_6
- \frac{9}{256} \zeta_5
- \frac{9}{1024} \zeta_4
- \frac{3}{1024} \zeta_3
- \frac{3}{16384}
\right) \epsilon^7
\right. \nonumber \\
&& \left. ~
+ \left(
\frac{3}{256} \zeta_3 \zeta_4
+ \frac{9}{256} \zeta_7
- \frac{15}{512} \zeta_6
- \frac{9}{1024} \zeta_5
- \frac{9}{4096} \zeta_4
- \frac{3}{512} \zeta_3^2
- \frac{3}{4096} \zeta_3
- \frac{3}{65536}
\right) \epsilon^8
\right) \frac{1}{N}
\nonumber \\
&&
+ \left( 
- \frac{3}{8} \epsilon
+ \frac{59}{64} \epsilon^2
- \frac{71}{256} \epsilon^3
- \left(
\frac{25}{128}
+ \frac{3}{8} \zeta_3
\right) \epsilon^4
+ \left(
 \frac{91}{128} \zeta_3
- \frac{19}{256}
- \frac{9}{32} \zeta_4
\right) \epsilon^5
\right. \nonumber \\
&& \left. ~~~~
+ \left(
\frac{273}{512} \zeta_4
- \frac{303}{16384}
- \frac{73}{512} \zeta_3
- \frac{21}{64} \zeta_5
\right) \epsilon^6
+ \left(
\frac{27}{65536}
+ \frac{305}{512} \zeta_5
- \frac{219}{2048} \zeta_4
- \frac{75}{256} \zeta_6
- \frac{63}{512} \zeta_3
- \frac{9}{64} \zeta_3^2
\right) \epsilon^7
\right. \nonumber \\
&& \left. ~~~~
+ \left(
\frac{123}{512} \zeta_3^2
+ \frac{535}{1024} \zeta_6
+ \frac{631}{131072}
- \frac{221}{2048} \zeta_5
- \frac{219}{4096} \zeta_3
- \frac{189}{2048} \zeta_4
- \frac{75}{256} \zeta_7
- \frac{27}{128} \zeta_3 \zeta_4
\right) \epsilon^8
\right) \frac{1}{N^2}
\nonumber \\
&&
+ \left( 
\frac{9}{16} \epsilon
- \frac{315}{128} \epsilon^2
+ \left(
\frac{527}{256}
- \frac{3}{8} \zeta_3
\right) \epsilon^3
+ \left(
 \frac{665}{256} \zeta_3
+ \frac{1809}{2048}
- \frac{9}{32} \zeta_4
\right) \epsilon^4
+ \left(
\frac{1995}{1024} \zeta_4
- \frac{5521}{1024} \zeta_3
- \frac{3661}{8192}
- \frac{39}{128} \zeta_5
\right) \epsilon^5
\right. \nonumber \\
&& \left. ~~~~
+ \left(
\frac{589}{256} \zeta_5
+ \frac{727}{256} \zeta_3
- \frac{16563}{4096} \zeta_4
- \frac{5817}{16384}
- \frac{135}{512} \zeta_6
- \frac{105}{256} \zeta_3^2
\right) \epsilon^6
\right. \nonumber \\
&& \left. ~~~~
+ \left(
\frac{1173}{512} \zeta_3^2
+ \frac{2181}{1024} \zeta_4
+ \frac{2661}{2048} \zeta_3
+ \frac{8455}{4096} \zeta_6
- \frac{23519}{131072}
- \frac{9943}{2048} \zeta_5
- \frac{507}{2048} \zeta_7
- \frac{315}{512} \zeta_3 \zeta_4
\right) \epsilon^7
\right. \nonumber \\
&& \left. ~~~~
+ \left(
\frac{27}{2560} \zeta_{5,3}
+ \frac{3519}{1024} \zeta_3 \zeta_4
+ \frac{7983}{8192} \zeta_4
+ \frac{17409}{8192} \zeta_7
+ \frac{20645}{8192} \zeta_5
- \frac{71825}{16384} \zeta_6
- \frac{18455}{65536} \zeta_3
- \frac{16011}{4096} \zeta_3^2
- \frac{5301}{10240} \zeta_8
\right. \right. \nonumber \\
&& \left. \left. ~~~~~~~~
-~ \frac{2385}{32768}
- \frac{315}{512} \zeta_3 \zeta_5
\right) \epsilon^8
\right) \frac{1}{N^3} ~+~ \order \left( \frac{1}{N^4} \right)
\end{eqnarray}
where the multiple zeta $\zeta_{5,3}$ first appears at $\order(\epsilon^8)$ in 
the third order large $N$ term. For the remaining exponents the analogous
expansions are
\begin{eqnarray}
\eta_\phi &=& \epsilon 
+ \left( 
- \frac{3}{2} \epsilon
+ \frac{7}{8} \epsilon^2
+ \frac{11}{32} \epsilon^3
+ \left(
\frac{19}{128}
- \frac{3}{8} \zeta_3
\right) \epsilon^4
+ \left(
\frac{7}{32} \zeta_3
+ \frac{35}{512}
- \frac{9}{32} \zeta_4
\right) \epsilon^5
+ \left(
\frac{11}{128} \zeta_3
+ \frac{21}{128} \zeta_4
+ \frac{67}{2048}
- \frac{9}{32} \zeta_5
\right) \epsilon^6
\right. \nonumber \\
&& \left. ~~~~~~
+ \left(
\frac{19}{512} \zeta_3
+ \frac{21}{128} \zeta_5
+ \frac{33}{512} \zeta_4
+ \frac{131}{8192}
- \frac{15}{64} \zeta_6
- \frac{3}{64} \zeta_3^2
\right) \epsilon^7
\right. \nonumber \\
&& \left. ~~~~~~
+ \left(
\frac{7}{256} \zeta_3^2
+ \frac{33}{512} \zeta_5
+ \frac{35}{256} \zeta_6
+ \frac{35}{2048} \zeta_3
+ \frac{57}{2048} \zeta_4
+ \frac{259}{32768}
- \frac{27}{128} \zeta_7
- \frac{9}{128} \zeta_3 \zeta_4
\right) \epsilon^8
\right) \frac{1}{N}
\nonumber \\
&&
+ \left( 
\frac{9}{4} \epsilon
- \frac{51}{32} \epsilon^2
- \left(
\frac{281}{128}
+ \frac{3}{2} \zeta_3
\right) \epsilon^3
+ \left(
\frac{9}{2} \zeta_3
+ \frac{13}{64}
- \frac{9}{8} \zeta_4
\right) \epsilon^4
+ \left(
\frac{27}{8} \zeta_4
+ \frac{33}{128}
- \frac{159}{64} \zeta_3
- \frac{3}{4} \zeta_5
\right) \epsilon^5
\right. \nonumber \\
&& \left. ~~~~
+ \left(
\frac{99}{32} \zeta_5
+ \frac{2127}{8192}
- \frac{477}{256} \zeta_4
- \frac{363}{256} \zeta_3
- \frac{15}{32} \zeta_6
- \frac{3}{4} \zeta_3^2
\right) \epsilon^6
\right. \nonumber \\
&& \left. ~~~~
+ \left(
\frac{11}{128} \zeta_3
+ \frac{63}{32} \zeta_3^2
+ \frac{315}{128} \zeta_6
+ \frac{7557}{32768}
- \frac{1089}{1024} \zeta_4
- \frac{537}{256} \zeta_5
- \frac{9}{8} \zeta_3 \zeta_4
- \frac{9}{32} \zeta_7
\right) \epsilon^7
\right. \nonumber \\
&& \left. ~~~~
+ \left(
\frac{33}{512} \zeta_4
+ \frac{189}{64} \zeta_3 \zeta_4
+ \frac{279}{128} \zeta_7
+ \frac{345}{2048} \zeta_3
+ \frac{12121}{65536}
- \frac{1075}{1024} \zeta_5
- \frac{945}{512} \zeta_6
- \frac{267}{256} \zeta_3^2
- \frac{21}{32} \zeta_8
\right. \right. \nonumber \\
&& \left. \left. ~~~~~~~~
-~ \frac{15}{16} \zeta_3 \zeta_5
\right) \epsilon^8
\right) \frac{1}{N^2} ~+~ \order \left( \frac{1}{N^3} \right)
\end{eqnarray}
\begin{eqnarray}
\eta_{\phi^2} &=&
\left( 
- 3 \epsilon
+ \frac{5}{2} \epsilon^2
+ \frac{1}{4} \epsilon^3
+ \left(
\frac{1}{8}
- \frac{3}{4} \zeta_3
\right) \epsilon^4
+ \left(
\frac{1}{16}
+ \frac{5}{8} \zeta_3
- \frac{9}{16} \zeta_4
\right) \epsilon^5
+ \left(
\frac{1}{16} \zeta_3
+ \frac{1}{32}
+ \frac{15}{32} \zeta_4
- \frac{9}{16} \zeta_5
\right) \epsilon^6
\right. \nonumber \\
&& \left. ~
+ \left(
\frac{1}{32} \zeta_3
+ \frac{1}{64}
+ \frac{3}{64} \zeta_4
+ \frac{15}{32} \zeta_5
- \frac{15}{32} \zeta_6
- \frac{3}{32} \zeta_3^2
\right) \epsilon^7
\right. \nonumber \\
&& \left. ~
+ \left(
\frac{1}{64} \zeta_3
+ \frac{1}{128}
+ \frac{3}{64} \zeta_5
+ \frac{3}{128} \zeta_4
+ \frac{5}{64} \zeta_3^2
+ \frac{25}{64} \zeta_6
- \frac{27}{64} \zeta_7
- \frac{9}{64} \zeta_3 \zeta_4
\right) \epsilon^8
\right) \frac{1}{N}
\nonumber \\
&&
+ \left( 
27 \epsilon
- \frac{147}{4} \epsilon^2
+ \left(
\frac{15}{16}
- \frac{51}{4} \zeta_3
\right) \epsilon^3
+ \left(
\frac{35}{4} \zeta_5
+ \frac{191}{64}
+ \frac{311}{8} \zeta_3
- \frac{153}{16} \zeta_4
\right) \epsilon^4
\right. \nonumber \\
&& \left. ~~~~
+ \left(
\frac{17}{8} \zeta_3^2
+ \frac{175}{16} \zeta_6
+ \frac{439}{256}
+ \frac{933}{32} \zeta_4
- \frac{507}{16} \zeta_3
- \frac{123}{4} \zeta_5
\right) \epsilon^5
\right. \nonumber \\
&& \left. ~~~~
+ \left(
\frac{19}{16} \zeta_3
+ \frac{51}{16} \zeta_3 \zeta_4
+ \frac{63}{4} \zeta_7
+ \frac{359}{8} \zeta_5
+ \frac{1495}{1024}
- \frac{2205}{64} \zeta_6
- \frac{1521}{64} \zeta_4
- \frac{183}{16} \zeta_3^2
\right) \epsilon^6
\right. \nonumber \\
&& \left. ~~~~
+ \left(
\frac{57}{64} \zeta_4
+ \frac{137}{16} \zeta_3 \zeta_5
+ \frac{149}{128} \zeta_3
+ \frac{625}{32} \zeta_3^2
+ \frac{4215}{4096}
+ \frac{4963}{256} \zeta_8
+ \frac{5625}{128} \zeta_6
- \frac{2929}{64} \zeta_7
- \frac{857}{32} \zeta_5
- \frac{549}{32} \zeta_3 \zeta_4
\right) \epsilon^7
\right. \nonumber \\
&& \left. ~~~~
+ \left(
\frac{75}{8} \zeta_3 \zeta_6
+ \frac{101}{96} \zeta_3^3
+ \frac{109}{128} \zeta_5
+ \frac{411}{64} \zeta_4 \zeta_5
+ \frac{435}{512} \zeta_3
+ \frac{447}{512} \zeta_4
+ \frac{1875}{64} \zeta_3 \zeta_4
+ \frac{4031}{192} \zeta_9
+ \frac{6505}{128} \zeta_7
+ \frac{10871}{16384}
\right. \right. \nonumber \\
&& \left. \left. ~~~~~~~~
-~ \frac{29771}{512} \zeta_8
- \frac{6035}{256} \zeta_6
- \frac{737}{64} \zeta_3^2
- \frac{121}{4} \zeta_3 \zeta_5
\right) \epsilon^8
\right) \frac{1}{N^2} ~+~ \order \left( \frac{1}{N^3} \right) ~.
\end{eqnarray}
Finally the two corrections to scaling exponents are
\begin{eqnarray}
\omega_+ &=&
\epsilon
+ \left( 
15 \epsilon
- \frac{53}{4} \epsilon^2
- \frac{13}{16} \epsilon^3
+ \left(
\frac{15}{4} \zeta_3
- \frac{29}{64}
\right) \epsilon^4
+ \left(
\frac{45}{16} \zeta_4
- \frac{61}{256}
- \frac{53}{16} \zeta_3
\right) \epsilon^5
+ \left(
\frac{45}{16} \zeta_5
- \frac{159}{64} \zeta_4
- \frac{125}{1024}
- \frac{13}{64} \zeta_3
\right) \epsilon^6
\right. \nonumber \\
&& \left. ~~~~~~
+ \left(
\frac{15}{32} \zeta_3^2
+ \frac{75}{32} \zeta_6
- \frac{253}{4096}
- \frac{159}{64} \zeta_5
- \frac{39}{256} \zeta_4
- \frac{29}{256} \zeta_3
\right) \epsilon^7
\right. \nonumber \\
&& \left. ~~~~~~
+ \left(
\frac{45}{64} \zeta_3 \zeta_4
+ \frac{135}{64} \zeta_7
- \frac{509}{16384}
- \frac{265}{128} \zeta_6
- \frac{87}{1024} \zeta_4
- \frac{61}{1024} \zeta_3
- \frac{53}{128} \zeta_3^2
- \frac{39}{256} \zeta_5
\right) \epsilon^8
\right) \frac{1}{N}
\nonumber \\
&&
+ \left( 
-~ \frac{315}{2} \epsilon
+ \frac{1845}{16} \epsilon^2
+ \left(
\frac{207}{2} \zeta_3
+ \frac{8767}{64}
\right) \epsilon^3
+ \left(
\frac{621}{8} \zeta_4
- \frac{1113}{4} \zeta_3
- \frac{105}{2} \zeta_5
- 49
\right) \epsilon^4
\right. \nonumber \\
&& \left. ~~~~
+ \left(
\frac{4965}{32} \zeta_3
+ 198 \zeta_5
- \frac{3339}{16} \zeta_4
- \frac{1347}{128}
- \frac{525}{8} \zeta_6
- \frac{51}{4} \zeta_3^2
\right) \epsilon^5
\right. \nonumber \\
&& \left. ~~~~
+ \left(
\frac{657}{8} \zeta_3^2
+ \frac{6885}{32} \zeta_6
+ \frac{8325}{128} \zeta_3
+ \frac{14895}{128} \zeta_4
- \frac{44505}{4096}
- \frac{4641}{16} \zeta_5
- \frac{189}{2} \zeta_7
- \frac{153}{8} \zeta_3 \zeta_4
\right) \epsilon^6
\right. \nonumber \\
&& \left. ~~~~
+ \left(
\frac{1971}{16} \zeta_3 \zeta_4
+ \frac{8949}{32} \zeta_7
+ \frac{16479}{128} \zeta_5
+ \frac{24975}{512} \zeta_4
- \frac{139683}{16384}
- \frac{14889}{128} \zeta_8
- \frac{2205}{8} \zeta_6
- \frac{561}{4} \zeta_3^2
\right. \right. \nonumber \\
&& \left. \left. ~~~~~~~~
-~ \frac{411}{8} \zeta_3 \zeta_5
- \frac{349}{16} \zeta_3
\right) \epsilon^7
\right. \nonumber \\
&& \left. ~~~~
+ \left(
\frac{1587}{8} \zeta_3 \zeta_5
+ \frac{8289}{128} \zeta_3^2
+ \frac{24917}{512} \zeta_5
+ \frac{28785}{256} \zeta_6
+ \frac{92337}{256} \zeta_8
- \frac{195131}{32768}
- \frac{4971}{16} \zeta_7
- \frac{4875}{1024} \zeta_3
- \frac{4031}{32} \zeta_9
\right. \right. \nonumber \\
&& \left. \left. ~~~~~~~~
-~ \frac{1683}{8} \zeta_3 \zeta_4
- \frac{1233}{32} \zeta_4 \zeta_5
- \frac{1047}{64} \zeta_4
- \frac{225}{4} \zeta_3 \zeta_6
- \frac{101}{16} \zeta_3^3
\right) \epsilon^8
\right) \frac{1}{N^2} ~+~ \order \left( \frac{1}{N^3} \right)
\end{eqnarray}
and
\begin{eqnarray}
\omega_- &=&
\epsilon
+ \left( 
- \frac{3}{2} \epsilon^2
+ \frac{7}{8} \epsilon^3
+ \frac{11}{32} \epsilon^4
+ \left(
\frac{19}{128}
- \frac{3}{8} \zeta_3
\right) \epsilon^5
+ \left(
\frac{7}{32} \zeta_3
+ \frac{35}{512}
- \frac{9}{32} \zeta_4
\right) \epsilon^6
\right. \nonumber \\
&& \left. ~~~~~~
+ \left(
\frac{11}{128} \zeta_3
+ \frac{21}{128} \zeta_4
+ \frac{67}{2048}
- \frac{9}{32} \zeta_5
\right) \epsilon^7
+ \left(
\frac{19}{512} \zeta_3
+ \frac{21}{128} \zeta_5
+ \frac{33}{512} \zeta_4
+ \frac{131}{8192}
- \frac{15}{64} \zeta_6
- \frac{3}{64} \zeta_3^2
\right) \epsilon^8
\right) \frac{1}{N}
\nonumber \\
&&
+ \left( 
\frac{87}{32} \epsilon^2
+ \left(
\frac{27}{8} \zeta_3
- \frac{293}{128}
\right) \epsilon^3
+ \left(
\frac{81}{32} \zeta_4
- \frac{419}{128}
- \frac{375}{32} \zeta_3
\right) \epsilon^4
+ \left(
\frac{13}{4} \zeta_5
+ \frac{1945}{1024}
+ \frac{2059}{128} \zeta_3
- \frac{1125}{128} \zeta_4
\right) \epsilon^5
\right. \nonumber \\
&& \left. ~~~~
+ \left(
\frac{49}{32} \zeta_3^2
+ \frac{385}{128} \zeta_6
+ \frac{6177}{512} \zeta_4
- \frac{4293}{512} \zeta_3
- \frac{2139}{8192}
- \frac{165}{16} \zeta_5
\right) \epsilon^6
\right. \nonumber \\
&& \left. ~~~~
+ \left(
\frac{147}{64} \zeta_3 \zeta_4
+ \frac{401}{128} \zeta_7
+ \frac{3425}{256} \zeta_5
+ \frac{24045}{32768}
- \frac{12879}{2048} \zeta_4
- \frac{4725}{512} \zeta_6
- \frac{2211}{2048} \zeta_3
- \frac{693}{128} \zeta_3^2
\right) \epsilon^7
\right. \nonumber \\
&& \left. ~~~~
+ \left(
\frac{85}{32} \zeta_3 \zeta_5
+ \frac{3701}{512} \zeta_3^2
+ \frac{4137}{1024} \zeta_8
+ \frac{9177}{8192} \zeta_3
+ \frac{23955}{2048} \zeta_6
- \frac{7059}{1024} \zeta_5
- \frac{6633}{8192} \zeta_4
- \frac{4773}{512} \zeta_7
- \frac{2079}{256} \zeta_3 \zeta_4
\right. \right. \nonumber \\
&& \left. \left. ~~~~~~~~
-~ \frac{955}{65536}
\right) \epsilon^8
\right) \frac{1}{N^2} ~+~ \order \left( \frac{1}{N^3} \right)
\end{eqnarray}
where the $\order(\epsilon^4)$ terms of $\omega_\pm$ were provided in 
\cite{Manashov:2017rrx}. We note that the leading terms of both $\omega_\pm$
are the same unlike their $N$~$=$~$2$ perturbative counterparts in 
(\ref{peromegplu}) and (\ref{peromegmin}). This is because while the leading 
$\epsilon$ term of (\ref{peromegplu}) is solely $\epsilon$ the coefficient of 
the corresponding term (\ref{peromegmin}) is $N$ dependent being in particular
$\frac{\sqrt{(4 N^2 + 132 N + 9)}}{(2 N + 3)}$ as can be seen in
\cite{Karkkainen:1993ef} after adjusting for the different $N$ convention. 
Clearly this tends to unity as $N$~$\to$~$\infty$.

\section{Gross-Neveu model critical exponents}
\label{appgnn2}

For completeness we record the four loop $O(N)$ GN critical exponents here in 
the same conventions as the GNY theory. From \cite{Gracey:2016mio} we have
\begin{eqnarray}
\eta_\psi &=& \frac{(4N-1)}{8(2N-1)^2} \epsilon^2 ~-~ 
\frac{(2N-3)(4N-1)}{16(2N-1)^3} \epsilon^3 ~+~ 
\frac{(4N-1)(16N^2-44N+25)}{128(2N-1)^4} \epsilon^4 ~+~ \order(\epsilon^5)
\nonumber \\
\eta_\phi &=& 2 ~-~ \frac{2N}{(2N-1)} \epsilon ~-~ 
\frac{(4N-1)}{4(2N-1)^2} \epsilon^2 ~+~ 
\frac{N(4N-1)}{4(2N-1)^3} \epsilon^3 \nonumber \\
&& +~ \frac{(4N-1)(4(8N^2+14N-21)\zeta_3-16N^2+36N+5)}{64(2N-1)^4} 
\epsilon^4 ~+~ \order(\epsilon^5) \nonumber \\
\frac{1}{\nu} &=& \epsilon ~-~ \frac{1}{2(2N-1)} \epsilon^2 ~-~ 
\frac{(4N-3)}{8(2N-1)^2} \epsilon^3 ~+~ 
\frac{((4N-1)(N+3)+3(22N-17)\zeta_3)}{8(2N-1)^3} \epsilon^4 ~+~ 
\order(\epsilon^5)
\end{eqnarray}
in $d$~$=$~$2$~$+$~$\epsilon$ dimensions. Numerically for $N$~$=$~$2$ these
equate to
\begin{eqnarray}
\eta_\psi &=& 0.097222 \epsilon^2 ~-~ 0.016204 \epsilon^3 ~+~ 
0.000675 \epsilon^4 ~+~ \order(\epsilon^5) \nonumber \\
\eta_\phi &=& 2.000000 ~-~ 1.333333 \epsilon ~-~ 0.194444 \epsilon^2 ~+~
0.129630 \epsilon^3 ~+~ 0.270765 \epsilon^4 ~+~ \order(\epsilon^5) \nonumber \\
\frac{1}{\nu} &=& 1.000000 \epsilon ~-~ 0.166667 \epsilon^2 ~-~ 
0.069444 \epsilon^3 ~+~ 0.612808 \epsilon^4 ~+~ \order(\epsilon^5) ~.
\label{expd2numn2}
\end{eqnarray}
The situation for $\eta_\phi$ and $1/\nu$ is the same as their four dimensional
counterparts in that the coefficients of the higher order terms in $\epsilon$
are not decreasing in contrast to $\eta_\psi$.
 
\bibliography{refs}

\end{document}